\documentclass[review]{elsarticle}

\usepackage{lineno,hyperref}
\usepackage{graphicx}
\usepackage{subfig}
\usepackage[linesnumbered,ruled,lined]{algorithm2e}
\usepackage{float}
\usepackage{tikz}
\usepackage{epstopdf}
\usepackage{amssymb}
\usepackage{amsmath}
\usepackage{booktabs}
\usepackage{array}
\usepackage{multirow}
\usetikzlibrary{arrows.meta}
\usepackage{natbib}

\journal{X. X. X}

\begin{document}

\begin{frontmatter}
		
		\title{Physics-driven Learning of the Steady Navier-Stokes Equations using Deep Convolutional Neural Networks}
		\author[Hao's address]{Hao Ma}
		\ead{hao.ma@tum.de}		
		\author[Yuxuan's address]{Yuxuan Zhang}
		\ead{uruz7@live.com}
		\author[Nils's address]{Nils Thuerey}
		\ead{nils.thuerey@tum.de}
		\author[Xiangyu's address]{Xiangyu Hu \corref{mycorrespondingauthor}}
		\cortext[mycorrespondingauthor]{Corresponding author. Tel.: +49 89 289 16152.}
		\ead{xiangyu.hu@tum.de}
		\author[Hao's address]{Oskar J. Haidn}
		\ead{oskar.haidn@tum.de}
		
		\address[Hao's address]{Department of Aerospace and Geodesy, Technical University of Munich, 85748 Garching, Germany}
		\address[Yuxuan's address]{Beijing Aerospace Propulsion Institute, 100076, Beijing, China}
		\address[Nils's address]{Technical University of Munich, 85748, Garching, Germany}
		\address[Xiangyu's address]{Department of Mechanical Engineering, Technical University of Munich, 85748 Garching, Germany}

\begin{abstract}
Recently, physics-driven deep learning methods have shown particular promise 
for the prediction of physical fields, especially  
to reduce the dependency on large amounts of pre-computed training data. 
In this work, we target the physics-driven learning of complex flow fields with high resolutions.
We propose the use of \emph{Convolutional neural networks} (CNN) based U-net architectures 
to efficiently represent and reconstruct the input and output fields, respectively. 
By introducing Navier-Stokes equations and boundary conditions into loss functions, 
the physics-driven CNN is designed to predict corresponding steady flow fields directly. 
In particular, this prevents many of the 
difficulties associated with approaches employing fully connected neural networks.
Several numerical experiments are conducted to investigate the behavior of the CNN approach, 
and the results indicate that a first-order accuracy has been achieved. 
Specifically for the case of a flow around a cylinder, 
different flow regimes can be learned and the adhered ``twin-vortices" are predicted correctly. 
The numerical results also show that the training for multiple cases is accelerated significantly, 
especially for the difficult cases at low Reynolds numbers, and
when limited reference solutions are used as supplementary learning targets.

\end{abstract}


\end{frontmatter}
%%%%%%%%%%%%%%%%%%%%%%%%%%%%%%%%%%%%%%%%%%%%%%%%%%%%%%%%%%%%%%%%%%%%%%%%%%%%%%%%

\clearpage

%%%%%%%%%%%%%%%%%%%%%%%%%%%%%%%%%%%%%%%%%%%%%%%%%%%%%%%%%%%%%%%%%%%%%%%%%%%%%%%%
\section{Introduction}\label{sec:Introduction}
%%%%%%%%%%%%%%%%%%%%%%%%%%%%%%%%%%%%%%%%%%%%%%%%%%%%%%%%%%%%%%%%%%%%%%%%%%%%%%%%

In some practical fluid mechanics problems such as real-time or frequent query analysis, a large number of solutions for different initial/boundary condition combinations are to be considered \cite{carr1994quantitative,ray2004swarm,sharma2014cfd}.
For the traditional discrete analysis, numerical simulations have to be conducted repeatedly and the computational cost quickly becomes overly expensive \cite{yu2019flowfield}. 
In contrast to classical computational methods, machine learning approaches, and especially the field of deep learning that employs \emph{Neural Networks} (NN), have demonstrated their capabilities 
to predict flow fields rapidly and accurately \cite{brunton2019machine,duraisamy2019turbulence,thuerey2020deep}.

The previous research on flow field prediction using NN is mainly focused on data-driven methods. 
Besides the indirect way using closure model \cite{ling2016reynolds,parish2016paradigm}, the field solution can also be directly obtained from the network trained with a large number of samples \cite{hesthaven2018non,ribeiro2020deepcfd}.
However, for complex flows in practical engineering problems, the training samples very often require extraction, pre-processing and may be hard to obtain \cite{bai2017data}. 
Some data-driven learning work utilizes \emph{Computational Fluid Dynamics} (CFD) approach to generate the data sets \cite{guo2016convolutional,bhatnagar2019prediction,ma2020supervised}, and it does not really solve the demand of avoiding the big computational cost of discrete methods.

In order to remedy the above-mentioned shortcomings, physics-driven methods are a relatively new development.
By providing physics information, NN are able to directly obtain the field solution with much less or even no training data. 
Based on \emph{Multi-layer Perceptron} (MLP) \cite{ciregan2012multi}, Raissi et al. designed a \emph{Physics Informed Neural Networks}(PINN). 
Due to the constraint of loss function employing \emph{Partial Differential Equations} (PDEs), the outputs gradually approach the ones obeying the physics laws \cite{raissi2018hidden,raissi2019physics,lu2019deepxde}.
Also, Sun et al. used MLP to predict fluid flows, in which a specific prior ansatz was devised to force the network satisfying the geometric boundary of a flow \cite{sun2020surrogate}. 
However, due to the full connectivity between the neurons, MLP suffer from extensive memory requirements and statistical inefficiencies  \cite{goodfellow2016deep}. 
Therefore, it is difficult to handle well the multi-dimensional learning space with high-resolution physics fields containing much more details.
Taking the highest resolution solution in Ref. \cite{sharma2018weakly} as an example, the MLP with one single temperature channel has over 1 million weights.
Considering more complex fluid dynamics problems requiring multiple feature channels, the weight count would increase even further.
One avenue for alleviating this problem is to employ the reduced-order modeling to compress and reconstruct the flow fields apart from training the network \cite{hesthaven2018non,wang2019non}. 
However, this operation not only is complicated but also may introduce additional errors from the projection onto reduced space \cite{rowley2004model}.  
In addition, MLP architecture by itself does not take into account the spatial structure of data. 
The data points in the learning domain irrespective of their distance are treated in a same way \cite{le2017automatic}.
However, the physics laws represented with PDEs are based on the localities of data points, which suggests that the capability of NN to reflect this spatial relationship can be very important, especially for the physics-driven methods which are constrained only by PDEs. 

On the other hand, \emph{Convolutional Neural Networks}(CNN) represent a specialized and well-established type of NN to tackle the aforementioned challenges \cite{krizhevsky2012imagenet}.
In previous research using data-driven methods, CNNs have presented a good performance to predict high-fidelity physics solutions. 
E.g., without an extra reduced-order modeling step, the CNN can directly compress and reconstruct high-fidelity flow fields with a series of convolutional calculations \cite{thuerey2020deep}. 
In physics-driven methods, CNNs succeeded in solving simple physics problems which obey a single PDE, such as Laplace equation \cite{sharma2018weakly} and Darcy's law \cite{zhu2019physics}, achieving high computational efficiency in capturing multi-scale features of the physics fields. 
Meanwhile, CNNs also show the capacity to learn spatial connections between the adjacent data points \cite{ma2020combined}, or the long-term control of fluids with physical losses \cite{holl2019}.

In this paper, based on the idea of physics constraints and a specific CNN architecture, we propose a \emph{Physics-driven Convolutional Neural Networks} (PD-CNN) method. 
With this method, \emph{Navier-Stokes} (N-S) equations and boundary conditions are introduced as a loss function that is discretized on the computational mesh in a controlled manner. 
The geometry of the object and other flow conditions are additionally embedded in the input layer.
To our knowledge, this is the first attempt that using complex, discrete PDE formulations to drive CNNs for 
predicting physics fields.

%%%%%%%%%%%%%%%%%%%%%%%%%%%%%%%%%%%%%%%%%%%%%%%%%%%%%%%%%%%%%%%%%%%%%%%%%%%%%%%%
\section{Methodology}\label{sec:Methodology}
%%%%%%%%%%%%%%%%%%%%%%%%%%%%%%%%%%%%%%%%%%%%%%%%%%%%%%%%%%%%%%%%%%%%%%%%%%%%%%%%

%%%%%%%%%%%%%%%%%%%%%%%%%%%%%%%%%%%%%%%%%%%%%%%%%%%%%%%%%%%% 

In this section, the CNN architecture used to compress input and to reconstruct the output are described.
Then the physics-driven learning framework for N-S equations is introduced, while an accelerating approach employing reference targets is presented at last.

%%%%%%%%%%%%%%%%%%%%%%%%%%%%%%%%%%%%%%%%%%%%%%%%%%%%%%%%%%%% 
\subsection{U-net architecture of CNN}

The U-net is a widely-used architecture of CNN which is first designed for biomedical image segmentation \cite{ronneberger2015u} and has previously been used for flow field reconstruction with data-driven learning \cite{thuerey2020deep}. 
In this paper, we modify it to suit physics-driven learning approaches. 

\begin{figure}[!h]
	\centering
	\includegraphics[width=1\textwidth]{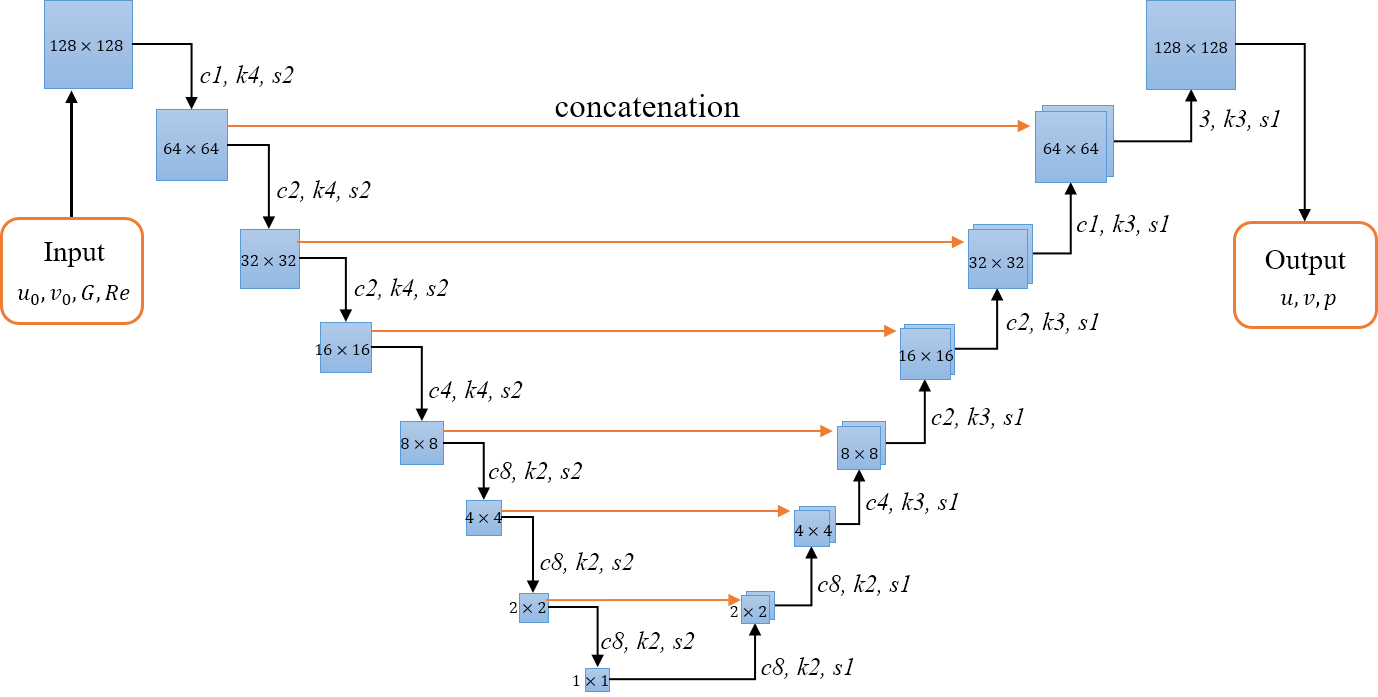}
	\caption{Schematic of U-net architecture. 
		Each blue box corresponds to a multi-channel feature map
		generated by a convolutional layer. 
		The resulting size are denoted in the box.
		Black corner arrows denote the down-sampling or up-sampling operation through convolutional layers.
		$cX, kX, sX$ shortly denotes channel factor $c=X$, convolutional kernel size $k=(X,X)$, stride $s=X$ respectively.
		And the channel number is the product of $c$ and a basic multiplier 64.
		The activation functions \emph{Rectified Linear Unit} (ReLU) is used to introduce the non-linearity. 
		Orange arrows denote ``skip-connections" via concatenation.}
	\label{figs:221-Unet}
\end{figure}

As Figure \ref{figs:221-Unet} shows, including the input and output layers, the U-net architecture consists of 17 layers and corresponding convolutional blocks.
The input layer consists of four channels. 
The first two, $u_{0}$ and $v_{0}$, are inflow velocities in both $x$ and $y$ directions, which are uniform non-dimensional values in the whole learning domain.
The geometry channel $G$ describes the shape of the object in the flow fields.
When there is an object in the flow, all values inside it be marked as 1 and the other as 0.
In this way, the geometry is embedded into the network and it is also used for evaluating physics loss as shown in later discussion.
To study the capability of the proposed method in characterizing the different patterns of flow, the Reynolds number is introduced as the fourth channel with the definition
\begin{equation}\label{con:mut}
Re=\frac{\rho v L}{\mu}
\end{equation}
where $\rho$ is density, $v$ inflow velocity, $L$ characteristic length, and $\mu$ the dynamic viscosity.
Since the $\rho$, $v$, $L$ are all unit values in this paper, $ Re $ is only depends on $\mu$.
The output layer consists of three channels $u$, $v$, and $p$,  which are velocities in both x and y directions and pressure respectively. These outputs are also non-dimensional values.

From inputs towards outputs, the network consists of two processes: encoding and decoding.
In the encoding process, the input fields are progressively down-sampled by convolutional calculations with corresponding kernels. 
In this way, the matrices with a size of $128 \times 128$ are reduced to a 512 component vector.
The decoding part works in an opposite manner, using an inverse convolutional process mirroring the behavior of the encoding part. 
Along with the increase of spatial resolution, the flow fields are reconstructed by up-sampling operations and convolutions. 

The core target of this work is to generate a flow field constrained by given physics laws. 
For steady problems, the PDEs, i.e. the mathematical expressions of underlying physics laws, represent the spatial relationships of adjacent positions. 
Similarly, the convolutional kernels extract the spacial feature of the receptive field consisting of a group of adjacent pixels. 
In the U-net architecture we used for our CNN, the encoding part is responsible for recognizing the geometry and inflow conditions of the flow field, in order to extract the necessary features representing the physics of the inputs using convolution operation layer by layer. 
These features are the basis for the subsequent decoding part. 
Here, the layers of the decoding process at different depths store the physical feature maps and the spacial relationship are recovered by the inverse convolutional calculation.
Eventually, the decoding part is able to reconstruct the proper flow field under the constrain of PDEs. 
In addition, there are the concatenations of the feature channels between encoding and decoding as the orange arrows denoted in Figure \ref{figs:221-Unet}. 
Duplicating the feature channels from the encoding blocks to the corresponding decoding ones, the ``skip connections" effectively double the number of feature channels in each decoding layer and enable the network to consider the information from the encoding layers, which extracted from the geometry and inflow conditions.

The architecture used in this paper is symmetrical, which means the encoding and decoding processes have the same depth, meanwhile, the amounts and dimensions of corresponding blocks are the same. 
However, the depths of the two processes are both adjustable. 
A coarse input compressed by fewer encoding layers is also able to generate a high-resolution solution reconstructed by more decoding layers. 
More details of the U-net architecture and convolutional block, including activation function, pooling, and dropout, can be found in Ref. \cite{thuerey2020deep} and \cite{paszke2019pytorch}.

%%%%%%%%%%%%%%%%%%%%%%%%%%%%%%%%%%%%%%%%%%%%%%%%%%%%%%%%%%%% 
\subsection{Physics-driven learning} 

The PDEs controlling the behavior of Newtonian fluid are Navier-Stokes (N-S) Equations. 
In our study, the steady and incompressible form of N-S Equations is chosen as follows:
\begin{equation}\label{con:NScontinuity}
	\nabla \cdot \textbf{\emph{U}}
	=0,
\end{equation}
\begin{equation}\label{con:NSmomentum}
	\textbf{\emph{U}} \cdot \nabla \textbf{\emph{U}}
	+\nabla \emph{P}
	-\mu\nabla^{2} \textbf{\emph{U}}
	=0,
\end{equation}
where $\textbf{\emph{U}}\equiv \textbf{\emph{U}}(u, v)$,  $\emph{P}$, $\mu$ are velocity, pressure and viscosity respectively. 
Equation (\ref{con:NScontinuity}) is the continuity equation, which imposes the incompressibilities of the fluid.  
Equation (\ref{con:NSmomentum}) is the momentum conservation equation, in which the first term represents the momentum convection, $\nabla \emph{P}$ the pressure gradient and $\mu\nabla \textbf{\emph{U}}$ the viscous dissipation.
\begin{figure}[!h]
	\centering
	\includegraphics[width=1\textwidth]{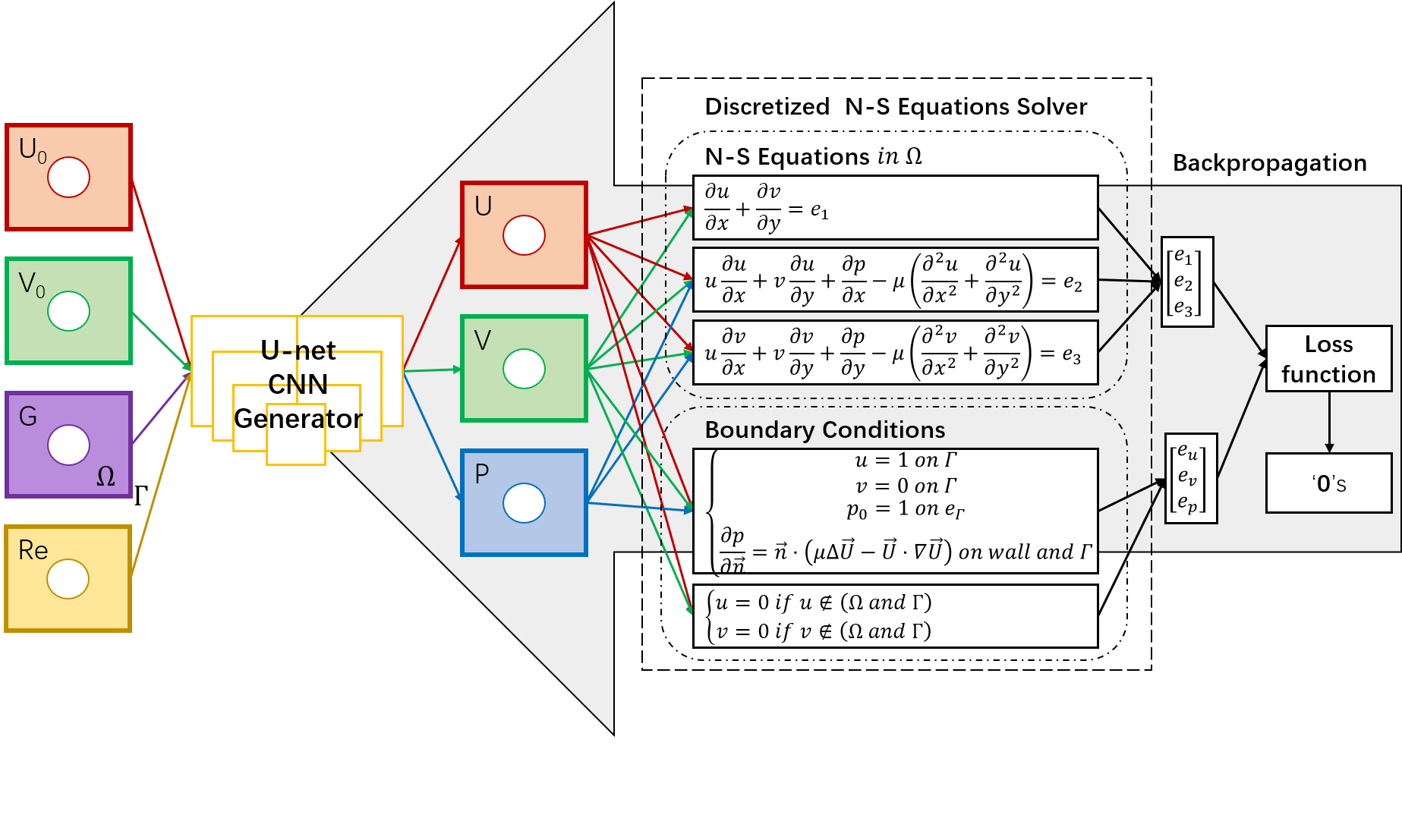}
	\caption{Physics-driven learning for solving the N-S equations. 
	The U-net CNN generates the solution. The backpropagation computes the gradient of the loss function and updates the weights of the CNN to satisfy the discretized N-S equations and boundary conditions.}
	\label{figs:231-Structure}
\end{figure}
As shown in Figure \ref{figs:231-Structure}, once the preliminary flow field is obtained from the U-net CNN generator, we apply the physical constraint to this field. 
The learning domain is separated as inner domain and boundaries, which are represented by $\Omega$ and $\Gamma $ respectively. 
In the inner domain $\Omega$, the left hand sides of N-S Equations are employed as loss function and 3 residuals can be obtained as
\begin{equation}\label{con:lossNS}
	\left\{
	\begin{array}{ccc}
		
		\frac{\partial u}{\partial x}+
		\frac{\partial v}{\partial y}=
		e_{1}    \\
		u\frac{\partial u}{\partial x}+
		v\frac{\partial u}{\partial y}+
		\frac{\partial p}{\partial x}-
		\mu\left(\frac{\partial^2 u}{\partial x^2}+\frac{\partial^2 u}{\partial y^2}\right)=
		e_{2}    \\
		u\frac{\partial v}{\partial x}+
		v\frac{\partial v}{\partial y}+
		\frac{\partial p}{\partial y}-
		\mu\left(\frac{\partial^2 v}{\partial x^2}+\frac{\partial^2 v}{\partial y^2}\right)=
		e_{3}    \\
		
	\end{array} \right.
\end{equation}
So the residuals of the N-S equations in $\Omega$ can be represented as
$E_{ \rm{\Omega} }=[
e_{1},
e_{2},
e_{3}       
]^{\rm T}$. 

%To construct the simple PDEs in algorithms, Ref. \cite{sharma2018weakly} proposed a unique convolutional filter, in which the weights are manually defined. While for the complex ones such as N-S equations, the weights have to be trained with a moderate number of samples. In order to remedy the restriction of data generation, we devised a type of convolutional filter as the partial differential operator to realize the central difference. 
In order to obtain differentiable formulation of the physics in the loss, we construct suitable 
convolutional filters to compute the N-S equations via finite differences. 
Similar approaches have been proposed previously 
for simple PDEs \cite{sharma2018weakly},
and the partial differential operators for two different dimensions are constructed separately \cite{long2018pde,long2019pde}. The construction via convolutions has the advantage that the backpropagation of a deep learning framework can be used, and the finite difference kernels yield well controlled accuracies for the derivative calculations.
Choosing the x direction as an example, the weights of the filters are represented as
\begin{equation}
	\rm{W}_{\frac{\partial}{\partial x}}
	={\left[ \begin{array}{ccc}
			0 & -0.5 & 0\\
			0 & 0 & 0\\
			0 & 0.5 & 0
		\end{array} 
		\right]},
	\rm{W}_{\frac{\partial^2}{\partial x^2}}
	={\left[ \begin{array}{ccc}
			0 & 1 & 0\\
			0 & -2 & 0\\
			0 & 1 & 0
		\end{array} 
		\right]}.
\end{equation}
After the rotating and moving operation through the matrix obtained from the last layer, the first- or second-order partial derivatives of local quantities are calculated. 
Choosing $u$ as an example, this procedure can be written as:
\begin{equation}
	g_{i,j}=\sum_{m=0}^{2}\sum_{n=0}^{2} u_{i+m-1,j+n-1} \cdot f_{m,n},
\end{equation}
where $ f_{m,n} $ is the convolutional filter and the $ g_{i,j} $ is the central difference of $u$ in each data point.

On the boundary $\Gamma$, including inflow and outflow side and walls, the Dirichlet and Neumann boundary conditions are considered as shown in \ref{figs:231-Structure}. 
The residuals of $u$, $v$ and $p$ are represented as 
$E_{\Gamma}=[
e_{u},    
e_{v},
e_{p}       
] ^{\rm T}$. 
Combining both as $E=[
E_{\Gamma},
E_{\Omega} ] ^{\rm T}$, the whole residual of the physics-driven method is obtained. 

To reduce the residuals, the CNN is trained in an iterative manner using a stochastic gradient descent variant (we employ Adam \cite{kingma2014adam}). 
After the CNN generator, the preliminary flow fields are introduced in the loss function, and then the residuals $E$ are obtained.
Once the backpropagation is applied, the weights and bias of CNN are adapted to minimize these physics residuals.
Eventually, the high-resolution flow fields which obey N-S Equations and corresponding boundary conditions can be obtained.

%%%%%%%%%%%%%%%%%%%%%%%%%%%%%%%%%%%%%%%%%%%%%%%%%%%%%%%%%%%% 
\subsection{Acceleration with reference targets} 

In the process of the physics-driven learning, the weights of CNN are adapted only to minimize the residuals of PDEs, the solution itself is not constrained, which means there is no target being offered for reference. 
In order to accelerate the convergence and eventually improve the training performance, besides the physical laws, we  provide additional reference targets for constraining the network.     

Similar to data-driven methods, there is a reference loss term comparing the difference between output and target, which is defined as
\begin{equation}\label{con:LossData}
	\mathcal{L}_{\rm{ref}}=\sum_{i=1}^{\mathcal{I}}\sum_{n=1}^{\mathcal{N}}  \left|  \mathcal{X}_{\rm{out}}-\mathcal{X}_{\rm{tar}}  \right|.
\end{equation}
The subscript ``ref'' here denotes reference targets. Generally, $\mathcal{X}_{\rm{out}}$ and $\mathcal{X}_{\rm{tar}} $ are output quantities and corresponding targets, respectively, $\mathcal{I} = targets\ amount$, meanwhile $\mathcal{N} = batch\ size$ denotes the amount of training data in one batch operation. 
The total loss considering both reference targets and physics laws can be represented as
\begin{equation}\label{con:weightedLossNew}
	\mathcal{L}=
	\mathcal{L}_{\rm{ref}}+
	{\rm R}*\mathcal{L}_{\rm{phy}},
\end{equation}
where $\mathcal{L}_{\rm{phy}}$ is the physics loss term considering the N-S equations and boundary conditions. R is a constant hyperparameter which is tuned to adapt the scales. With this weighted loss function, the different loss terms can be easily scaled to an equivalent magnitude.

\begin{figure}[!h]
	\centering
	\includegraphics[width=0.9\textwidth]{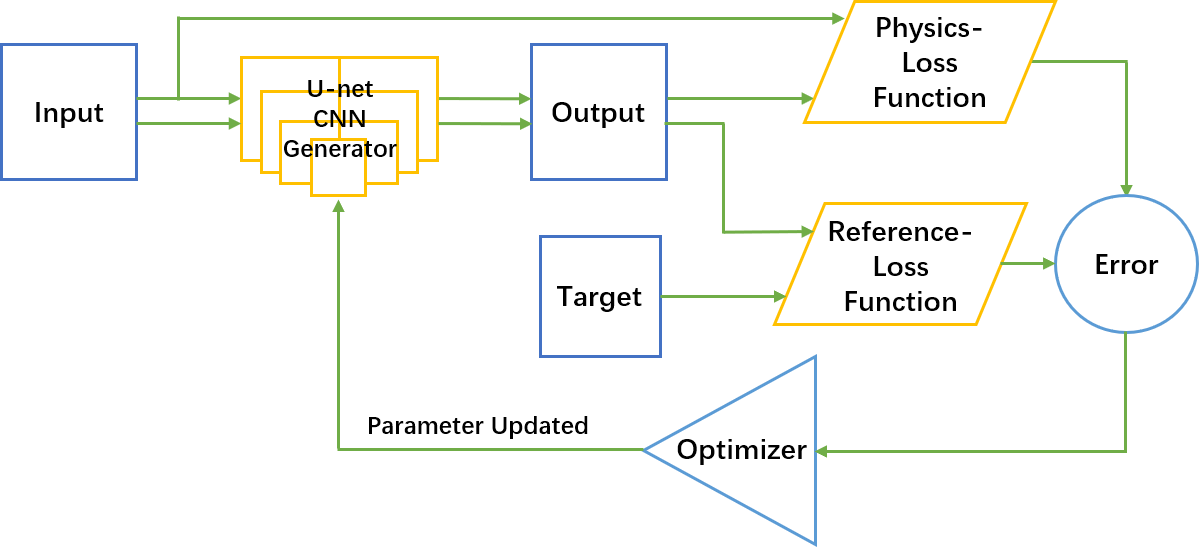}
	\caption{Acceleration with reference targets. One batch consists of the physics and reference groups. The outputs of physics group are introduced to physics loss function, while  the outputs of reference group are introduced to the reference loss function.}
	\label{figs:241-Combined}
\end{figure}

In the physics-driven training, a certain amount of randomly picked Reynolds numbers are input as one batch in each iterative step.
In contrast, in the accelerating approach with reference targets, we also use some constant Reynolds numbers besides the variable ones.
So, the new batch includes two groups as shown in Figure
\ref{figs:241-Combined}.
The cases in the random group vary in every iterative step and are trained with physics loss $ \mathcal{L}_{\rm{phy}} $. While the ones in constant group are fixed in the whole iterative process and are trained with reference loss $ \mathcal{L}_{\rm{ref}} $. 

This accelerating approach using reference targets is merely an enhancement of the original physics-driven method, which means the PD-CNN without references is sufficient to predict the final solution of the flow field. In order to clearly represent this property, section \ref{sec:Results} only presents the physics-driven-alone results while the enhancement of this reference acceleration will be discussed in section \ref{sec:Discussion}.

%%%%%%%%%%%%%%%%%%%%%%%%%%%%%%%%%%%%%%%%%%%%%%%%%%%%%%%%%%%%%%%%%%%%%%%%%%%%%%%%
\section{Results}\label{sec:Results}
%%%%%%%%%%%%%%%%%%%%%%%%%%%%%%%%%%%%%%%%%%%%%%%%%%%%%%%%%%%%%%%%%%%%%%%%%%%%%%%%

%%%%%%%%%%%%%%%%%%%%%%%%%%%%%%%%%%%%%%%%%%%%%%%%%%%%%%%%%%%% 
\subsection{Overview} 
In this section, several numerical experiments are conducted to estimate the capability of the proposed PD-CNN framework to predict steady laminar flow fields. 
The experiments follow two distinct patterns, single case and multiple cases. 

Single case means the PD-CNN only predicts the solution of one specific flow field. 
Given one specific combination of boundary conditions and fluid properties, the network is trained and then fixed to generate the unique corresponding solution. 
This pattern is similar to a traditional CFD simulation or a normal PINN training, and it is used to estimate the capacity of PD-CNN as an alternative to solve a specific problem.

Multiple cases mean the PD-CNN is used to obtain the solutions of multiple cases with a unique network.
The input inflow conditions are varied in a moderate range in the training stage. 
After that, given any of parameter combinations inside the span, the fixed network is able to generate the corresponding flow field. 
We use this pattern to estimate the capability of the PD-CNN learning different physics and whether these physics can coexist within one unique network.
Compared with CFD simulation, this capability allows the trained network to directly generate the solutions under different conditions with tiny computational cost, which makes the real-time or many-query analysis feasible.

In the study of single case, we choose several different 2D cases, such as Couette flow, Poiseuille flow, and flow around a cylinder. While for multiple cases, there are only cases of flow around cylinder. 
All of them are steady-state flows with low Reynolds numbers. 
For the Couette and Poiseuille flow, analytical solutions are used as references. For flow around cylinder cases, the numerical results calculated by the mature \emph{Finite Volume Method} (FVM) are used as references \cite{jasak2007openfoam}. 
The mesh for simulations is generated by Gmesh \cite{geuzaine2009gmsh} and an unstructured grid. Then the numerical solutions are interpolated into a Cartesian grid to compare with the PD-CNN results.

%%%%%%%%%%%%%%%%%%%%%%%%%%%%%%%%%%%%%%%%%%%%%%%%%%%%%%%%%%%% 
\subsection{Single case} 
\subsubsection{Couette flow}
\begin{figure}[!h]
	\centering
	\subfloat[Schematic]{
		\includegraphics[width=0.5\textwidth]{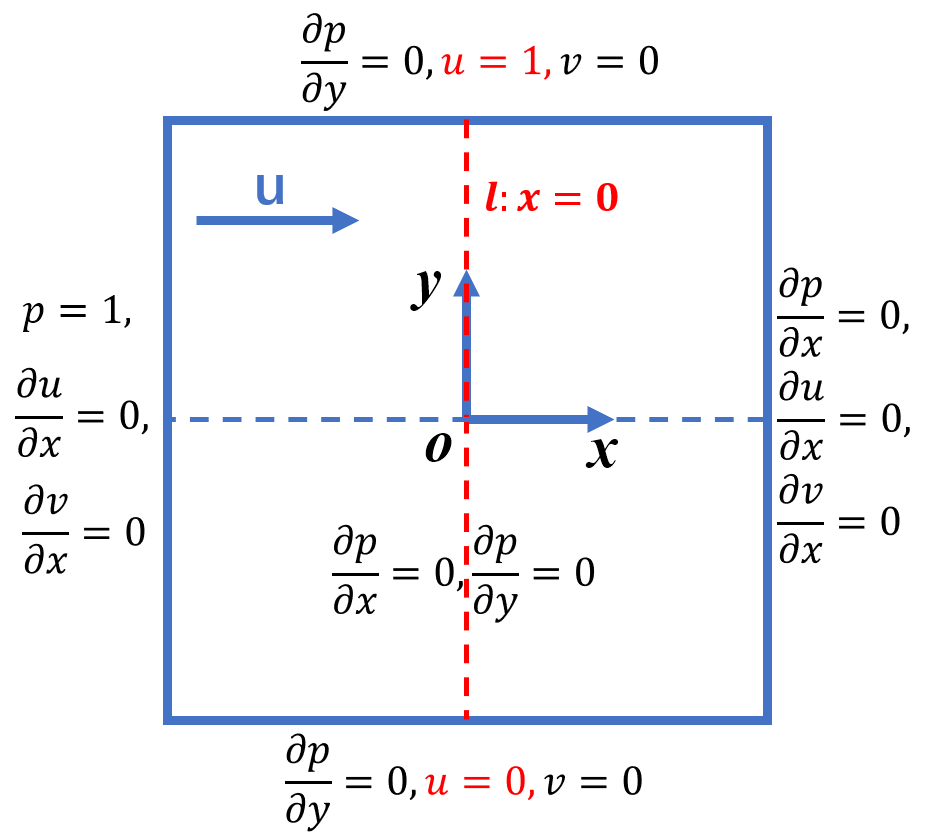}}
	\quad
	\subfloat[learning results on $l$]{
		\includegraphics[width=0.5\textwidth]{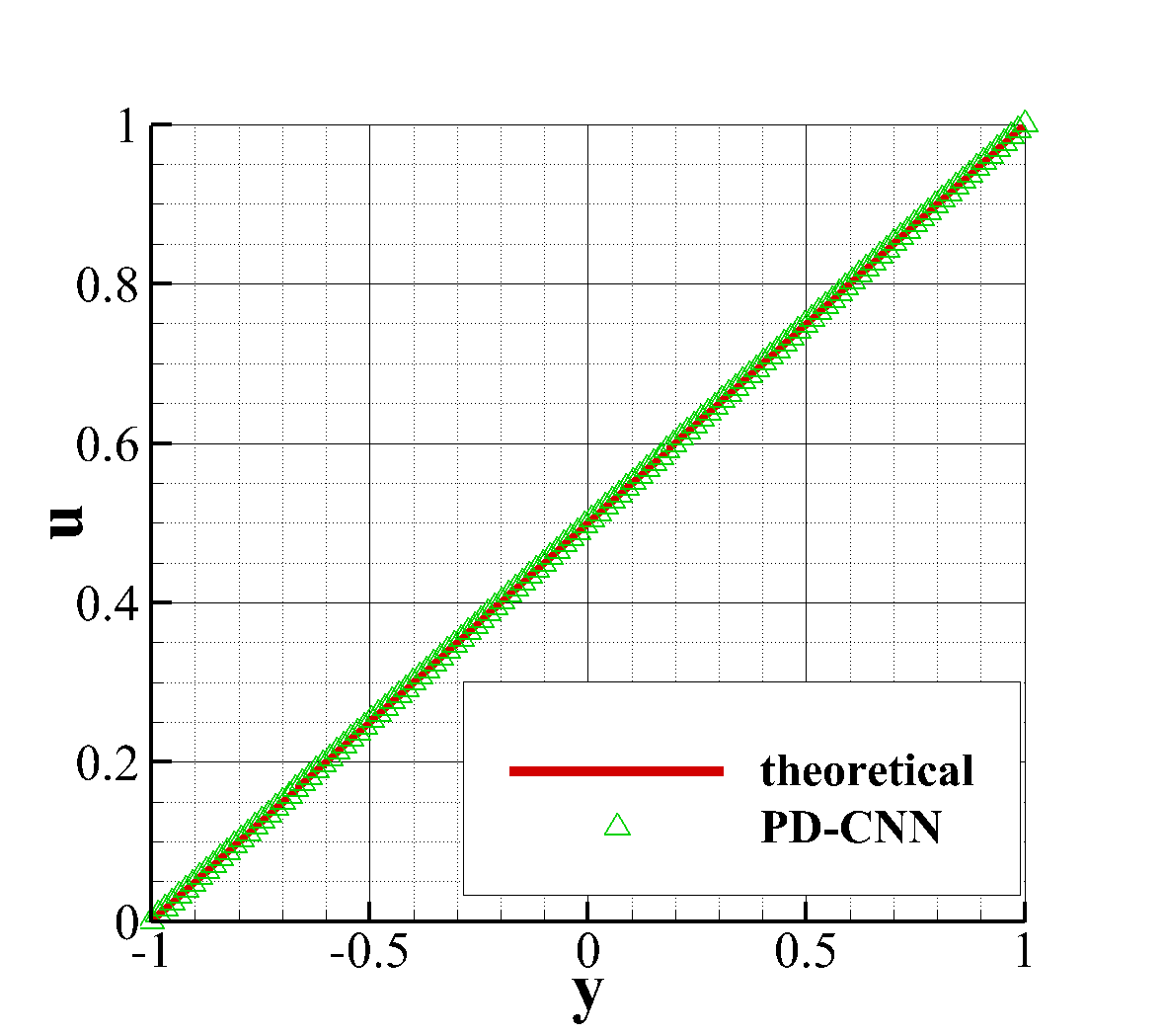}}
	\subfloat[learning results in the whole domain]{
		\includegraphics[width=0.5\textwidth]{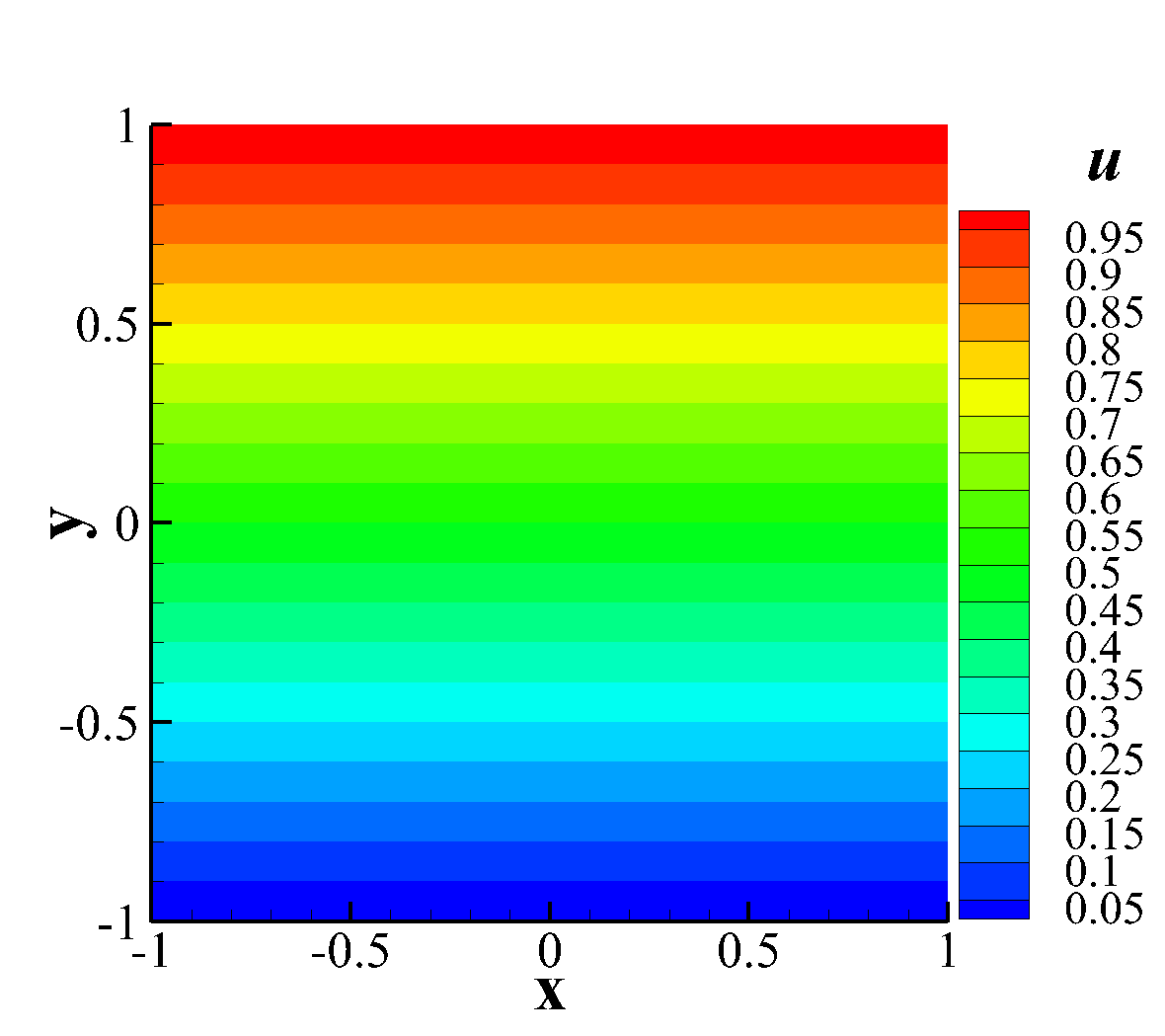}}
	\caption{Couette Flow.The learning domain is $x\in[-1,1]$ and $y\in[-1,1]$. The speed of top surface is $u=1$. And the viscosity coefficient $\mu=1$. For detecting line $l:x=0$, the angle between velocity profile line and horizontal axis is almost exactly $45^\circ$, which is the same as theoretical value. For the whole domain, $u$ along x-axis is isotropic which match the zero-gradient condition of $p$. Furthermore, there is no discontinuity on the boundary.}
	\label{figs:311-Couette}
\end{figure}

Couette flow is the flow of a viscous fluid in the space between two surfaces moving relatively, which is frequently used in engineering courses to illustrate shear-driven fluid motion. 
In our case, the top surface is moving along the positive direction of x-axis with a constant speed, and the bottom surface keeps still as shown in Figure \ref{figs:311-Couette}a. The results on the detecting line  $l:x=0$ and the entire domain are shown in \ref{figs:311-Couette}b and \ref{figs:311-Couette}c respectively.

The Couette flow is a tangential velocity-driven flow. The interference from $p$ and $v$ are relatively low. 
The only term that actually affects the flow is ${\partial^2 u}/{\partial y^2}$, which is the linear term of N-S Equations. 
Thus, this case tests the ability of PD-CNN to learn linear processes. It also means that a Laplace-like equation can be learned by the PD-CNN.

\subsubsection{Poiseuille flow}
\begin{figure}[!h]
	\centering
	\subfloat[Schematic]{
		\includegraphics[width=0.5\textwidth]{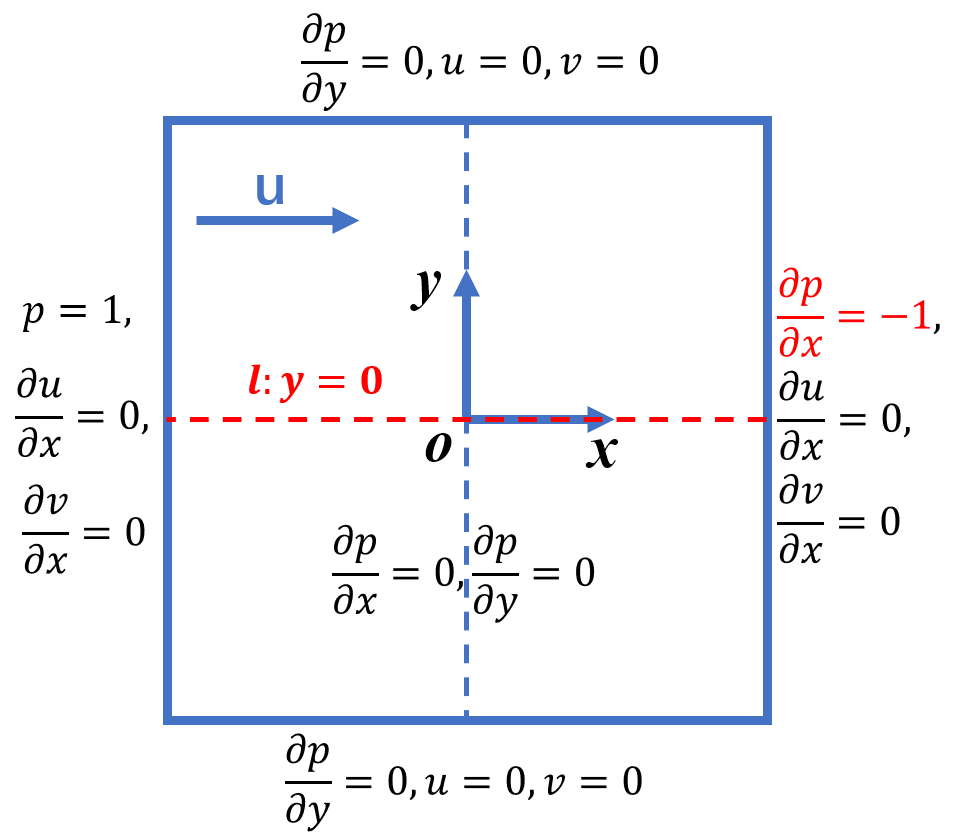}}
	\quad
	\subfloat[learning results on $l$]{
		\includegraphics[width=0.5\textwidth]{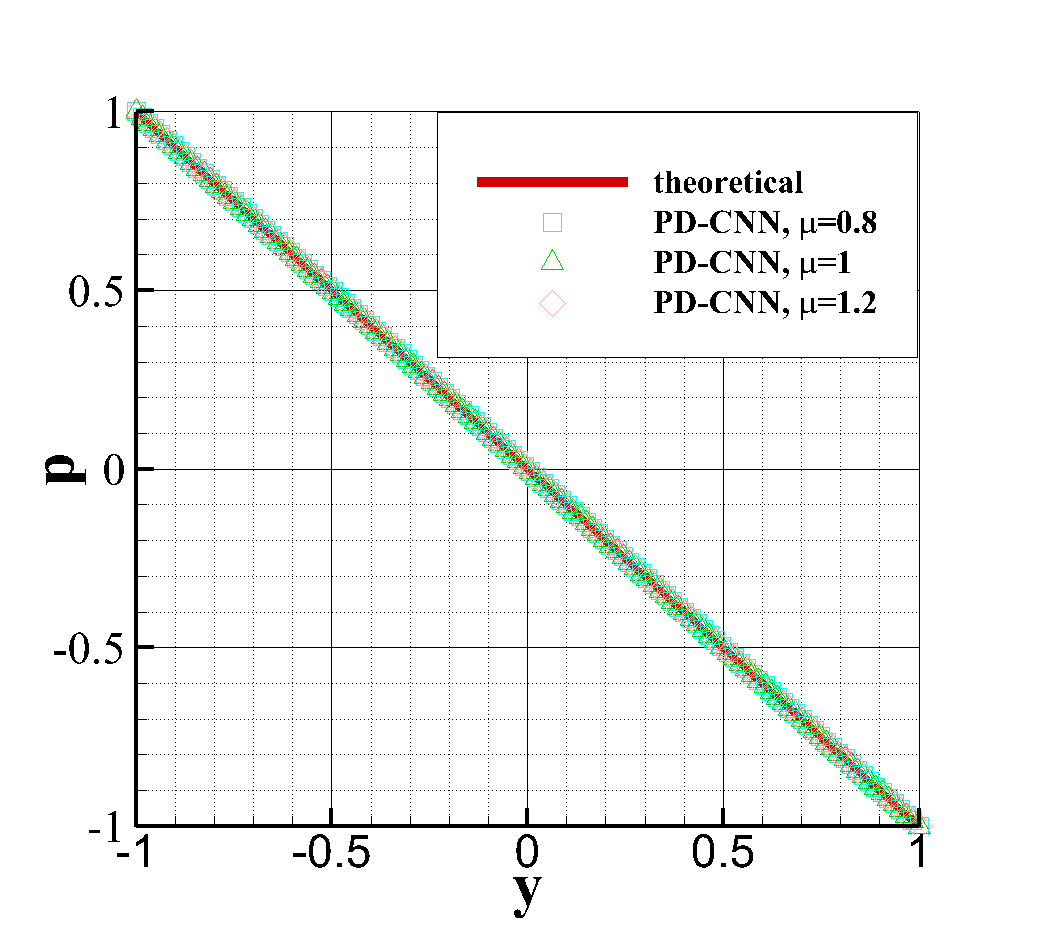}}
	\subfloat[learning results in the whole domain. $\mu=1$.]{
		\includegraphics[width=0.5\textwidth]{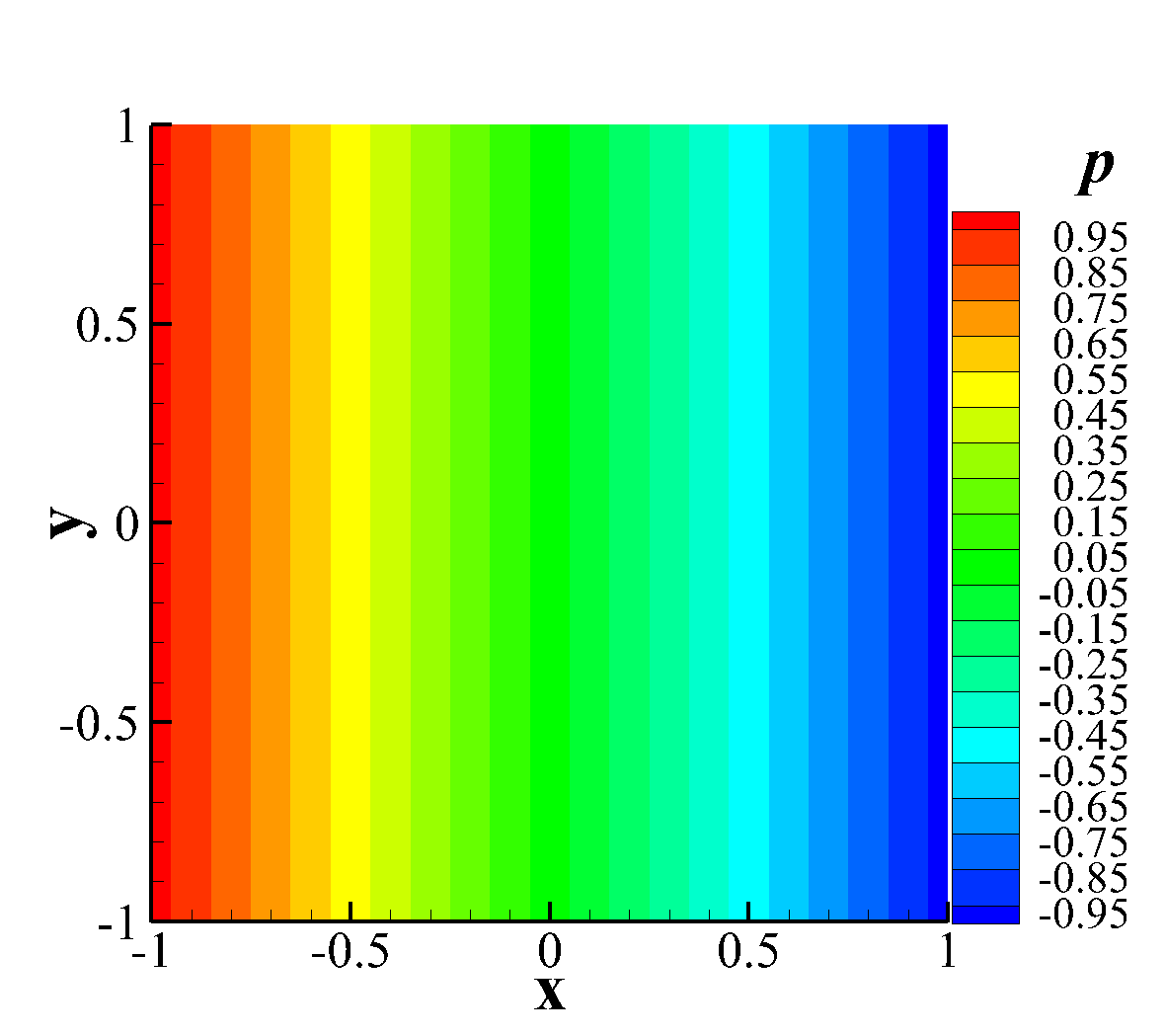}}
	\caption{Poiseuille Flow. The learning domain is  $x\in[-1,1]$ and $y\in[-1,1]$. Both top and bottom wall are stationary walls. For detecting line $l:y=0$, for all different $\mu$, the angle between velocity profile line and horizontal axis are almost exactly $135^\circ$, which is the same as theoretical values. For the whole domain, $p$ along y-axis is isotropic and there is no discontinuity on the boundary.}
	\label{figs:312-Poiseuille}
\end{figure}

A pressure-driven incompressible flow between two surfaces is calculated in this section. 
The boundary conditions are shown in Figure \ref{figs:312-Poiseuille}a. 
In the entire field, a gradient of pressure is imposed, which is the driven force of the movement of flow. For Poiseuille flow, the flow fields with different $\mu$ are the same, which is presented very well as shown in Figure \ref{figs:312-Poiseuille}b. This property indicates the rationality that designing multiple cases by altering $\mu $ in sub-section \ref{sec:MultipleCases}.

Although Poiseuille flow is also a linear flow, the flow is induced by the pressure gradient. The properties of all three variables, $u$, $v$, and $p$, are well represented. 
And the obtained flow field is swirl-free and symmetric. 
This proves that our treatment of the physics field is isotropic, and the error is effectively controlled.

\subsubsection{Flow around cylinder}
\begin{figure}[!h]
	\centering
	\includegraphics[width=0.5\textwidth]{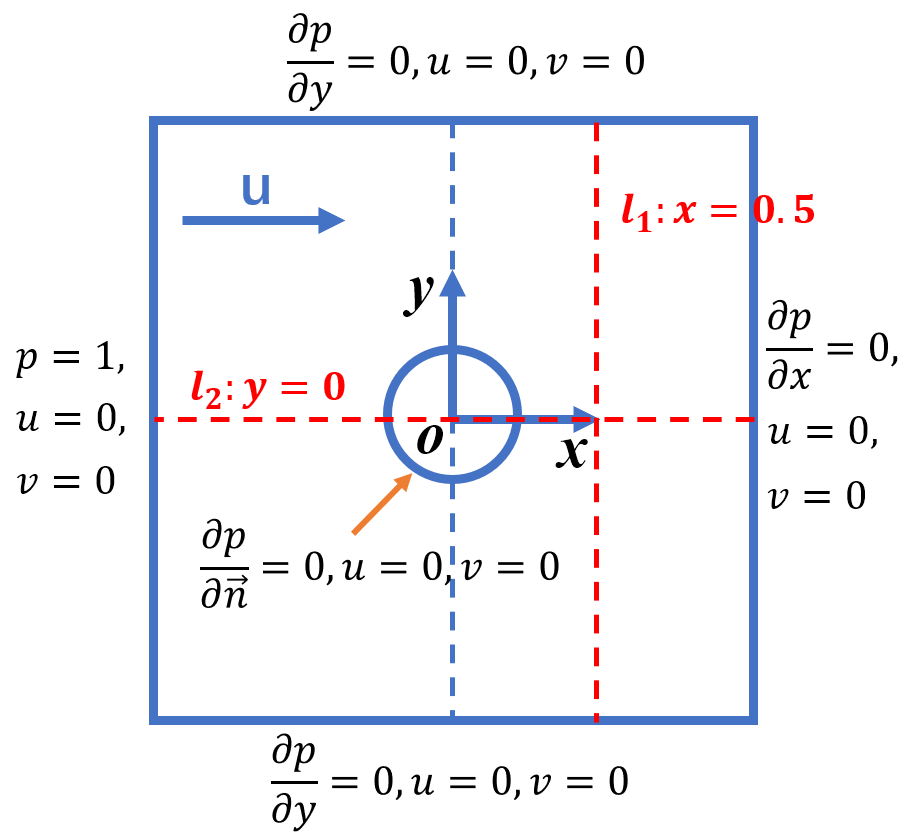}
	\caption{Schematic of flow around cylinder. The computing domain is $x\in[-1,1]$ and $y\in[-1,1]$ and diameter of the central cylinder is 0.3. There are two detecting lines $l_{1}$ and $l_{2}$ for later discussion.}
	\label{figs:313-FlowCylinder}
\end{figure}

In this section, the network is trained to solve a flow field around a cylinder(as shown in Figure \ref{figs:313-FlowCylinder}). 
Flow around a cylinder is a classic problem in fluid mechanics and its flow field exhibits different physical phenomena under different Reynolds numbers. 
Here, we consider solving the steady incompressible flows with Reynolds number from 1 to 20.
According to Ref. \cite{zdravkovich1998flow}, this range covers the two different regimes which are designated as creeping laminar state (L1) and laminar flow with steady separation (L2). 
In general, when the Reynolds number is smaller than approximately 5, the flow field will be the creeping laminar flow. 
Once the Reynolds number is larger than approximately 5, the flow field will transfer to a steady flow with a pair of symmetric trapped vortexes behind the cylinder. 
These two types of flow regimes are very different.

\begin{figure}[!h]
	\centering
	\subfloat[$p$]{
		\includegraphics[width=0.33\textwidth]{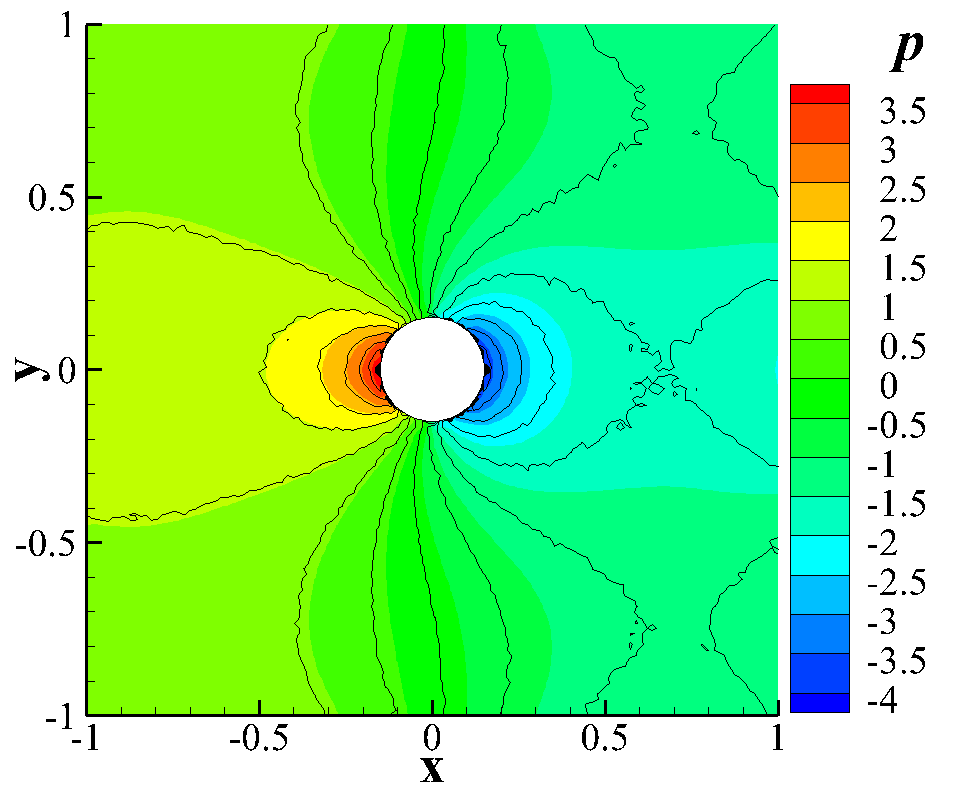}}
	\subfloat[$u$]{
		\includegraphics[width=0.33\textwidth]{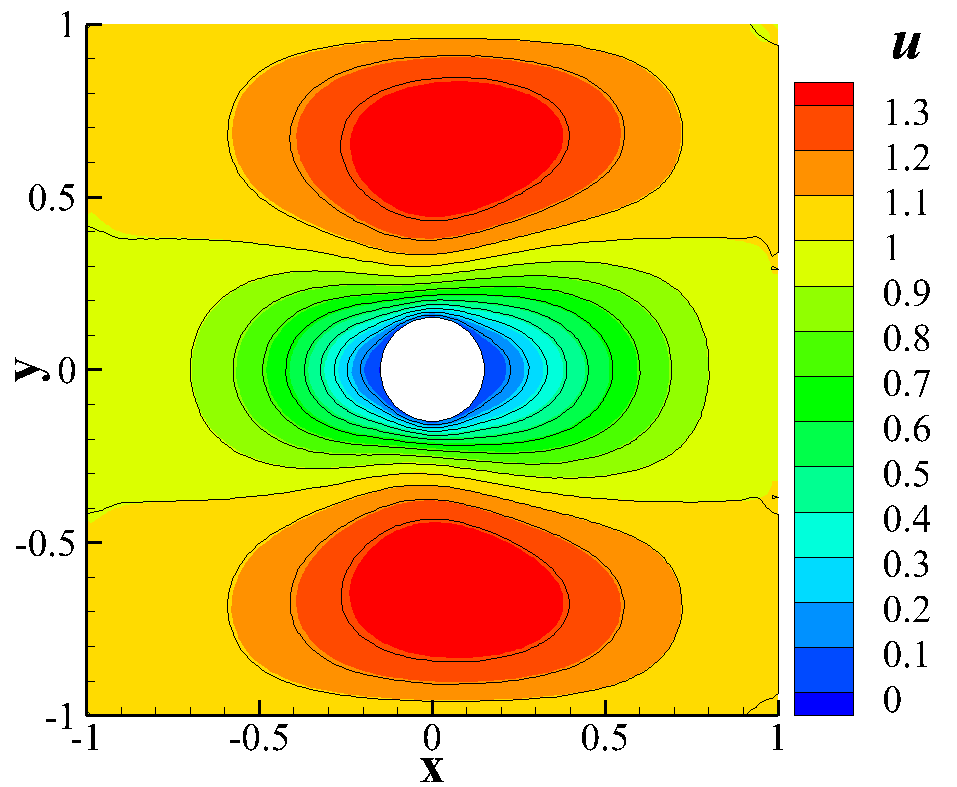}}
	\subfloat[$v$]{
		\includegraphics[width=0.33\textwidth]{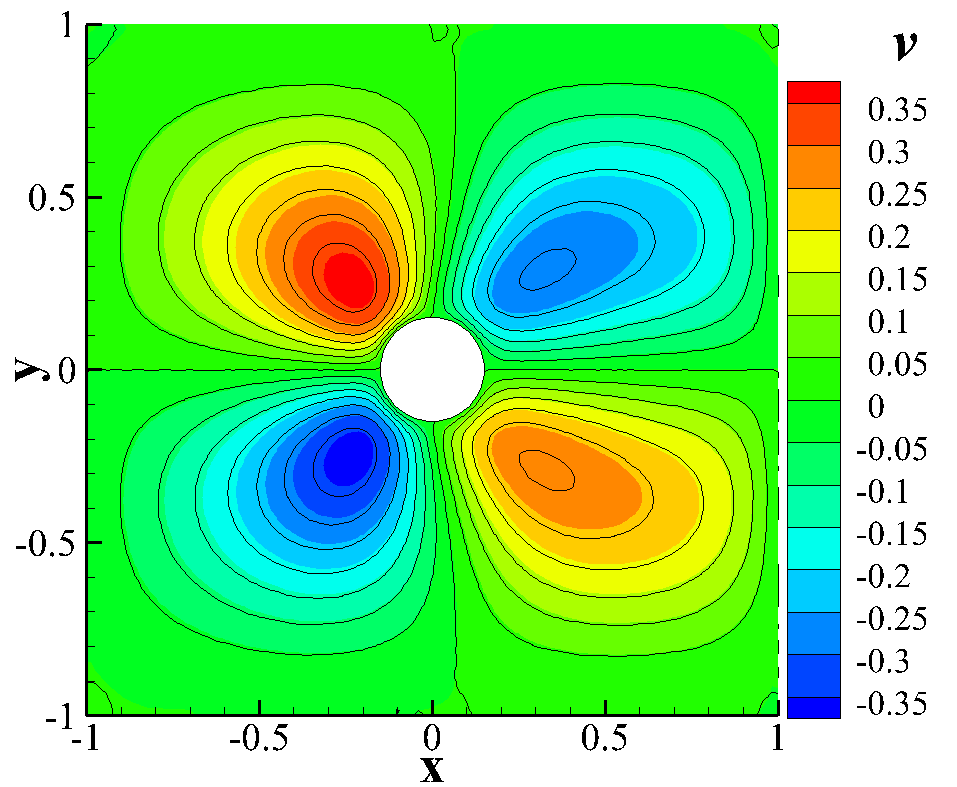}}
	\caption{Single case when $Re=1$. Black contour lines are FVM results, colorful contour flooding is PD-CNN results. For velocities, the learning results agree reference quite well. While for pressure, especially in the region behind the cylinder, the learning results have a slight deviation.}
	\label{figs:314-SingleRe1}
\end{figure}

\begin{figure}[!h]
	\centering
	\subfloat[$p$]{
		\includegraphics[width=0.33\textwidth]{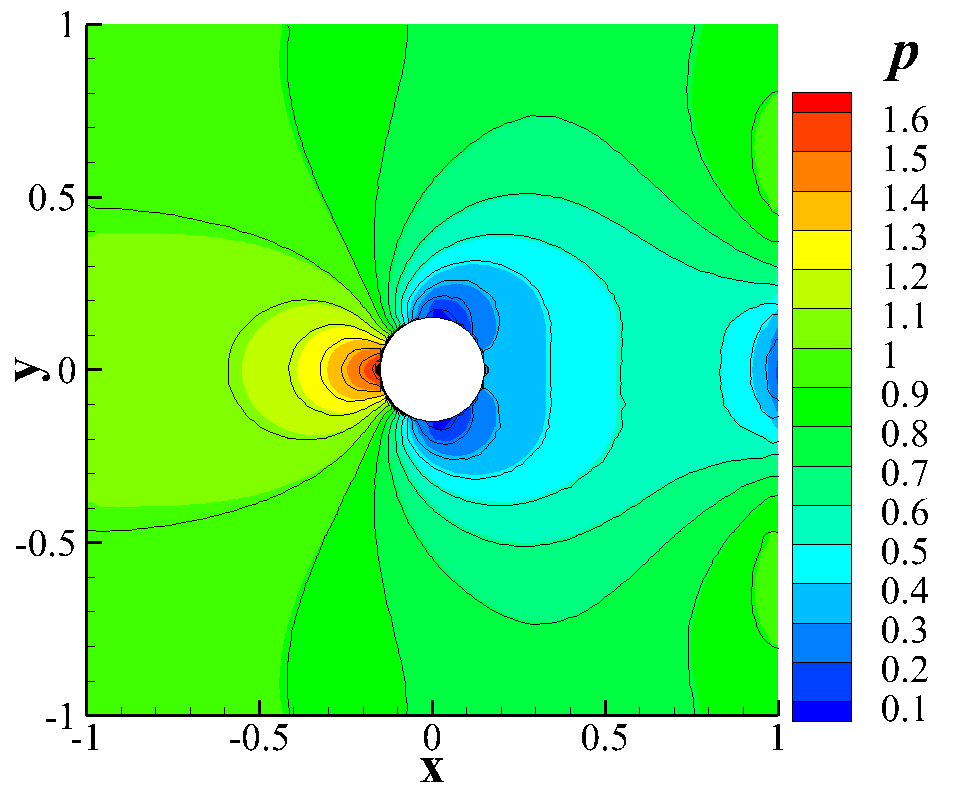}}
	\subfloat[$u$]{
		\includegraphics[width=0.33\textwidth]{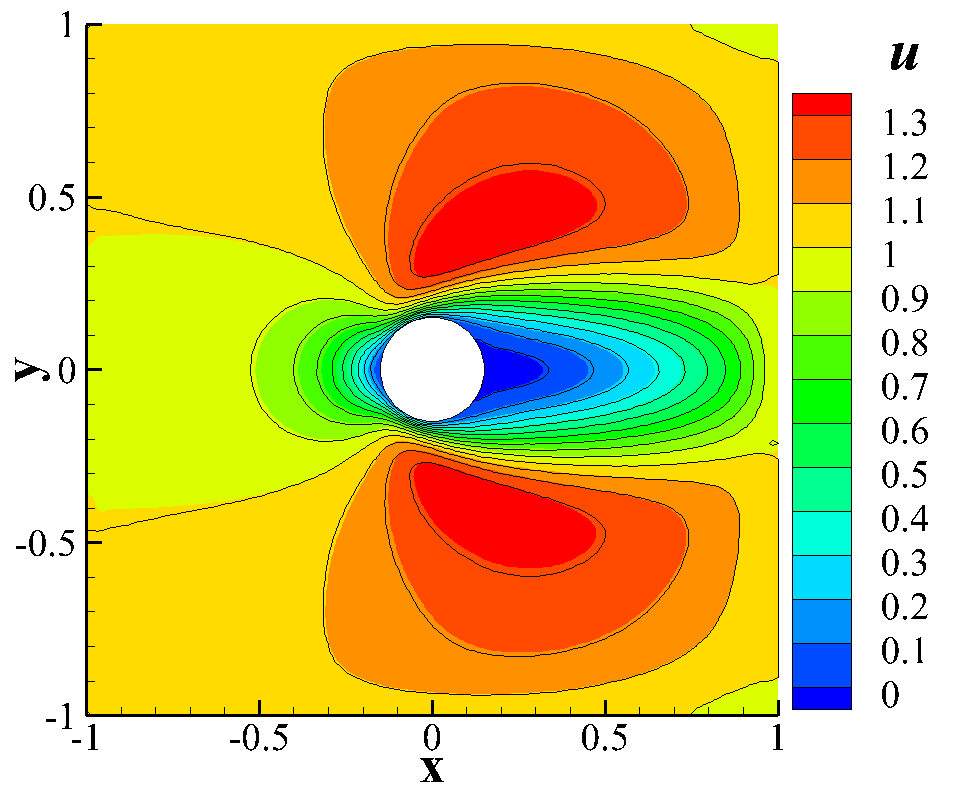}}
	\subfloat[$v$]{
		\includegraphics[width=0.33\textwidth]{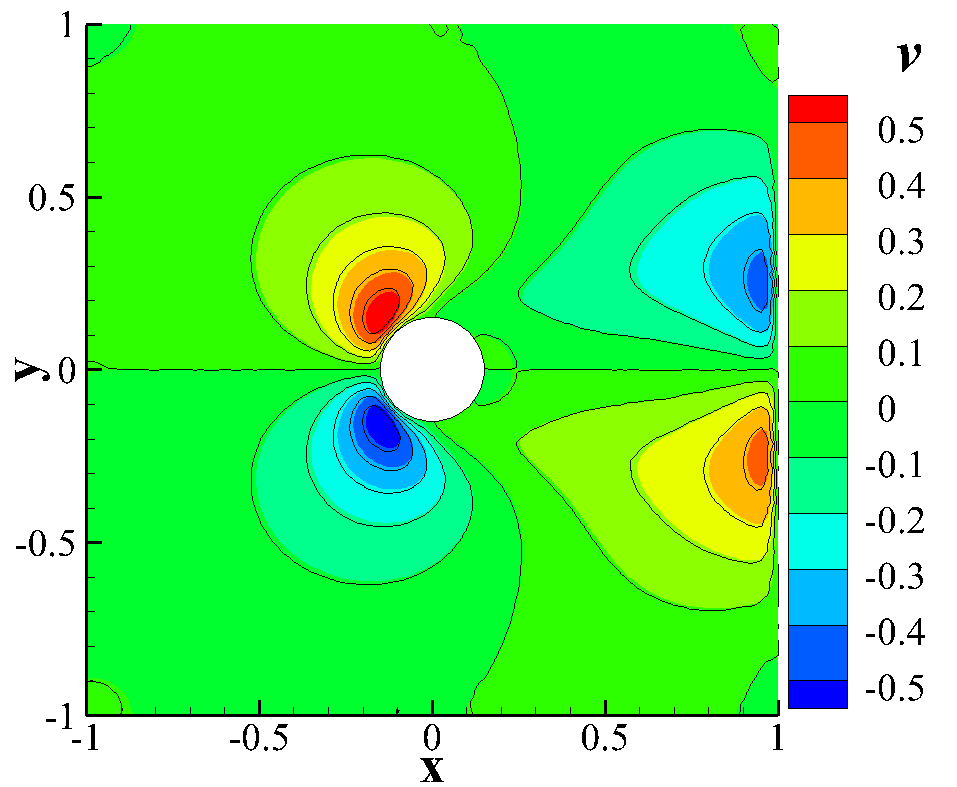}}
	\caption{Single case when $Re=20$. Black contour lines are FVM results, colorful contour flooding is PD-CNN results. For all velocities and pressure, the learning results agree reference quite well.}
	\label{figs:315-SingleRe20}
\end{figure}

Figure \ref{figs:314-SingleRe1} and Figure \ref{figs:315-SingleRe20} show the flow fields predicted by PD-CNN with $Re=1$ and $Re=20$ respectively. The learning results agree the numerical results obtained by FVM well.

%%%%%%%%%%%%%%%%%%%%%%%%%%%%%%%%%%%%%%%%%%%%%%%%%%%%%%%%%%%% 
\subsection{Multiple cases} \label{sec:MultipleCases}

For multiple cases training about flow around a cylinder, the CNN is trained with the variation of Reynolds number while the U-net architecture and inflow conditions are fixed. 
The Reynolds numbers, which are represented as the reciprocal of dynamic viscosity, are randomly picked between 0.8 and 20.2 in the whole training process.
It is similar to the generation of a large training set in data-driven methods. 
The random variation of the Reynolds number provides the PD-CNN inexhaustible training cases and forces the networks to consider the whole span.

\begin{figure}[!h]
	\centering
	\subfloat[$Re=1$]{
		\includegraphics[width=0.5\textwidth]{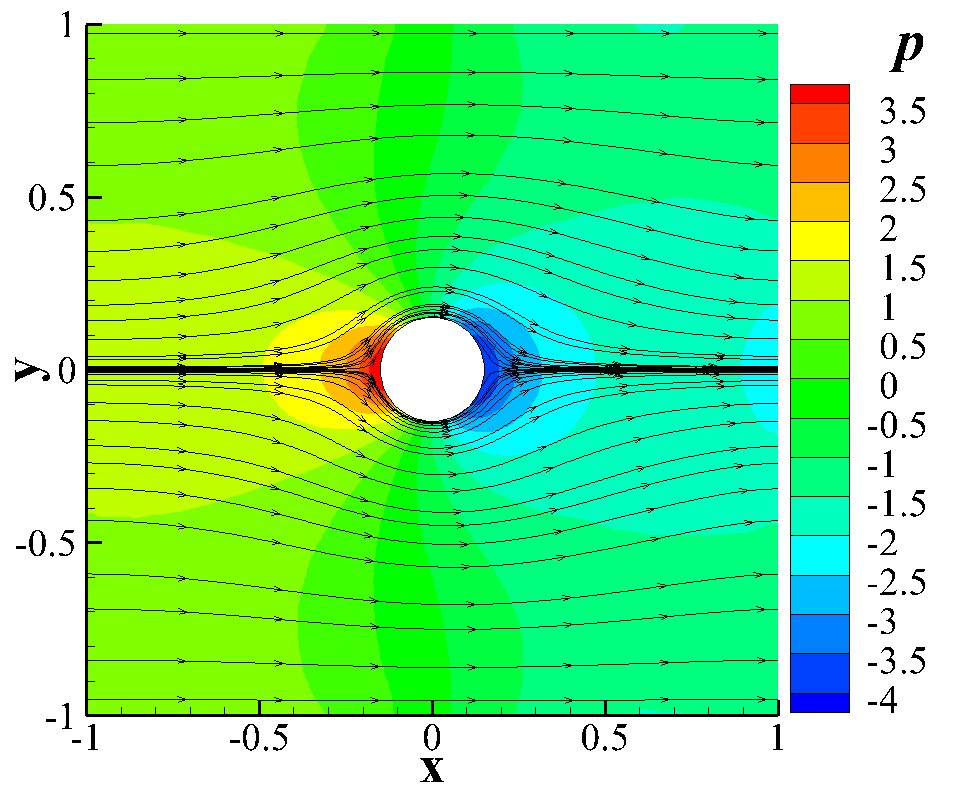}}
	\subfloat[$Re=5$]{
		\includegraphics[width=0.5\textwidth]{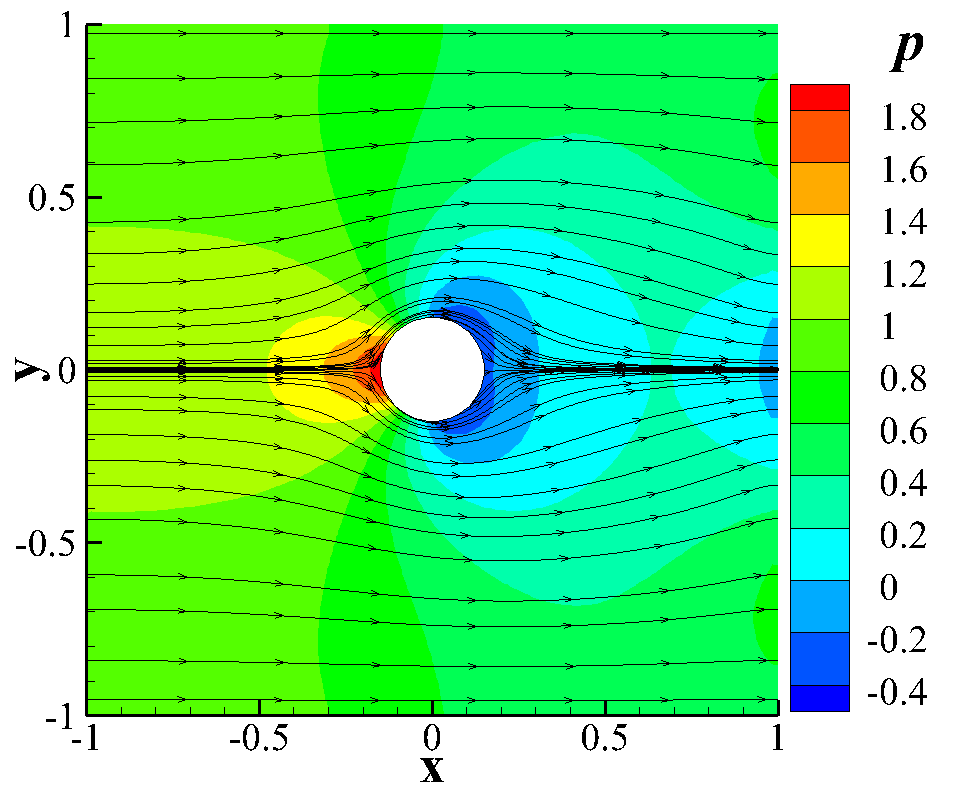}}
	\quad
	\subfloat[$Re=10$]{
		\includegraphics[width=0.5\textwidth]{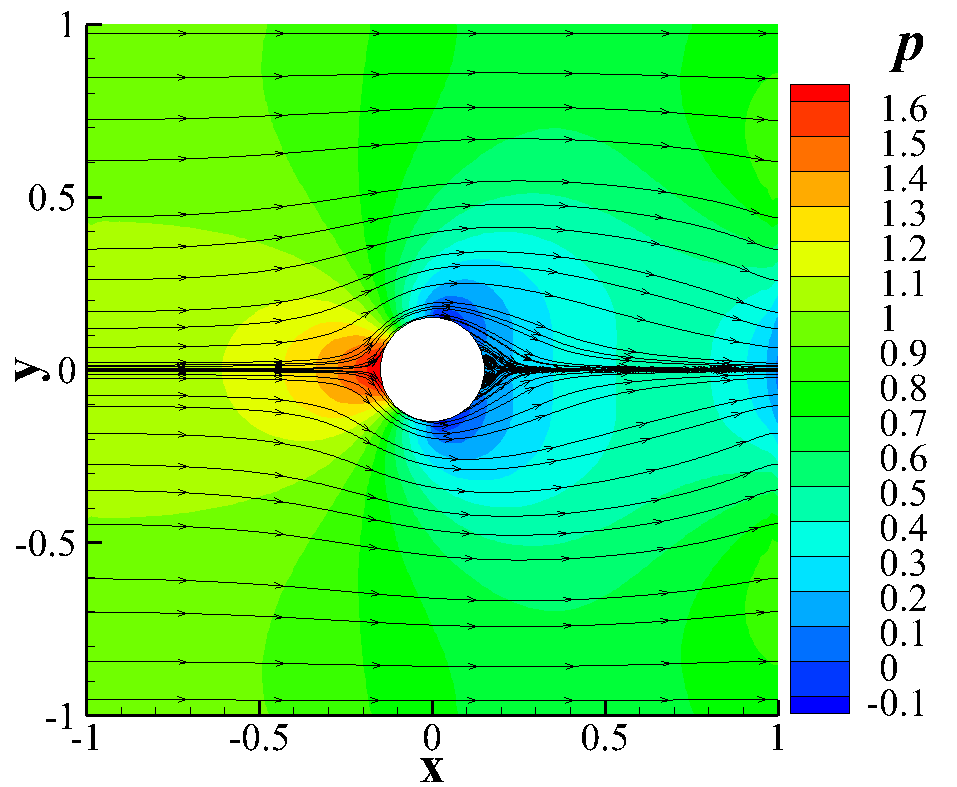}}
	\subfloat[$Re=20$]{
		\includegraphics[width=0.5\textwidth]{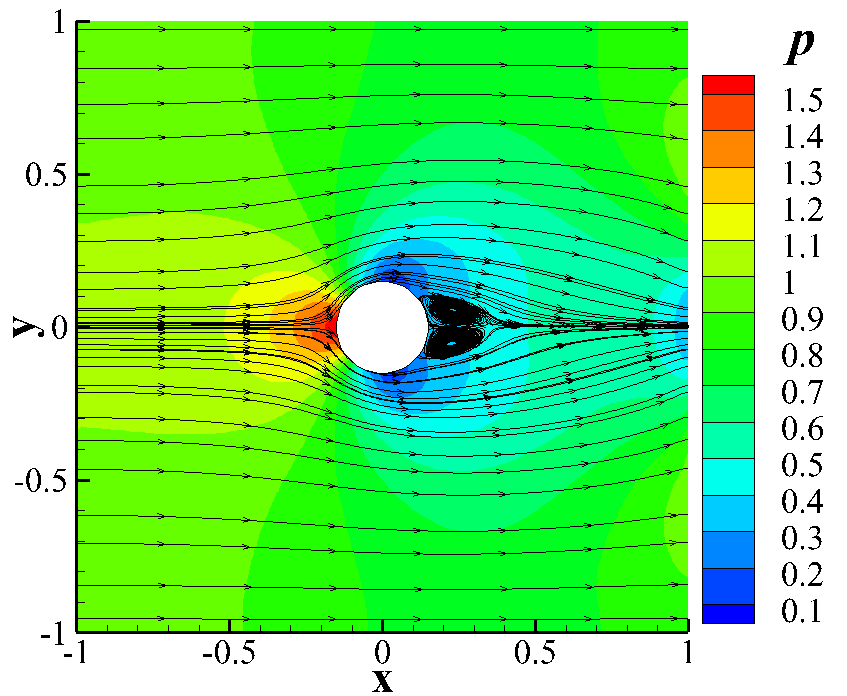}}
	\caption{Stream Traces with different $Re$. The results discriminate two different flow regimes and represent the evolution of the vortices along with the growing of $Re$. 
	After the creeping flow regime, (a) and (b), a pair of symmetric contra-rotating vortices appears in the rear of the cylinder, (c) and (d).}
	\label{figs:321-StreamTraces}
\end{figure}

Figure \ref{figs:321-StreamTraces} shows the capability of PD-CNN to reconstruct the two flow regimes L1 and L2.
After the L1 regime, the flow separates on the cylinder surface and the wake behind started to form the contra-rotating vortices. 
In the whole range of L2, the pair of symmetric vortices, adhere stably behind the cylinder. 
For the relatively small value of the Reynolds number, the closed wake region which contains the ``twin-vortices'' is presented clearly in the rear of the cylinder.
And with the Reynolds number increasing, this pair of vortices grows and becomes more and more elongated in the flow direction. 
The results talked above show a unique network obtained by PD-CNN is able to learn the different physical properties and predict the distinct flow fields according to input parameters.

The contra-rotating vortices can be noticed from the stream trace pictures of $Re=8$ but not $Re=7$. 
But it can not be concluded that the critical Reynolds number at which the pair of symmetric contra-rotating vortices begin to form is between 7 and 8. 
Because in the neighborhood of the critical Reynolds number, the dimension of the contra-rotating vortices is very small and can not be directly observed in the stream trace pictures.
By employing the method proposed by Taneda \cite{taneda1956experimental}, the critical Reynolds number can be estimated. 
\begin{figure}[!htbp]
	\centering
	\includegraphics[width=0.8\textwidth]{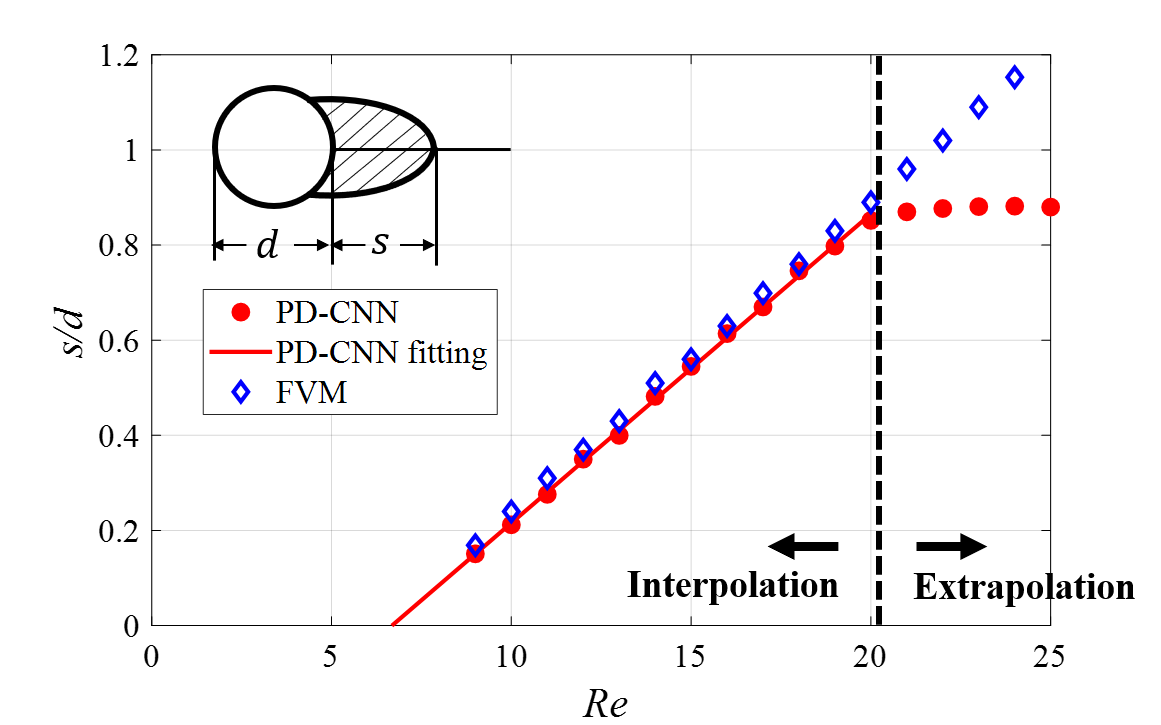}
	\caption{Size of the vortices pair against $Re$. By linear fitting, the critical Reynolds number at which the vortices pair begin to form in the rear of a cylinder is obtained.}
	\label{figs:322-VorticesLength}
\end{figure}  
As shown in Figure \ref{figs:322-VorticesLength}, the sizes of the twin-vortices are measured against the Reynolds numbers. 
It can be observed that as the Reynolds number increases, the length of twin-vortices grows. 
Hence, from the linear curve fit the critical Reynolds number can be deduced. 
So the pair of the contra-rotating vortices begin to form in the rear of the cylinder at $Re=6.63$, which is close with the FVM result, $Re=6.1$ \cite{rajani2009numerical}.

When $Re<20.2$, reasonably good agreements are observed between the learning results and the numerical solutions for the size of the closed wake.
We also evaluate the extrapolation capabilities of the network with the Reynolds numbers between 20.2 and 25.
However, as the comparison shows, the predicted length of the twin-vortices has bigger deviations and the extrapolated flow fields don't represent reliable references. 
It can be concluded that for the PD-CNN it is hard to accurately recover the neighboring flow fields which are completely missing at the learning stage. 
This phenomenon reflects the fundamentally interpretive characteristic of NN modeling and the model is merely well approximated in the training span \cite{brunton2019machine}.
If the objective field is relatively complex and the trainable parameters of NN are extensive, the network deteriorates the extrapolation accuracy badly.  

\begin{figure}[!h]
	\centering
	\subfloat[$p$]{
		\includegraphics[width=0.5\textwidth]{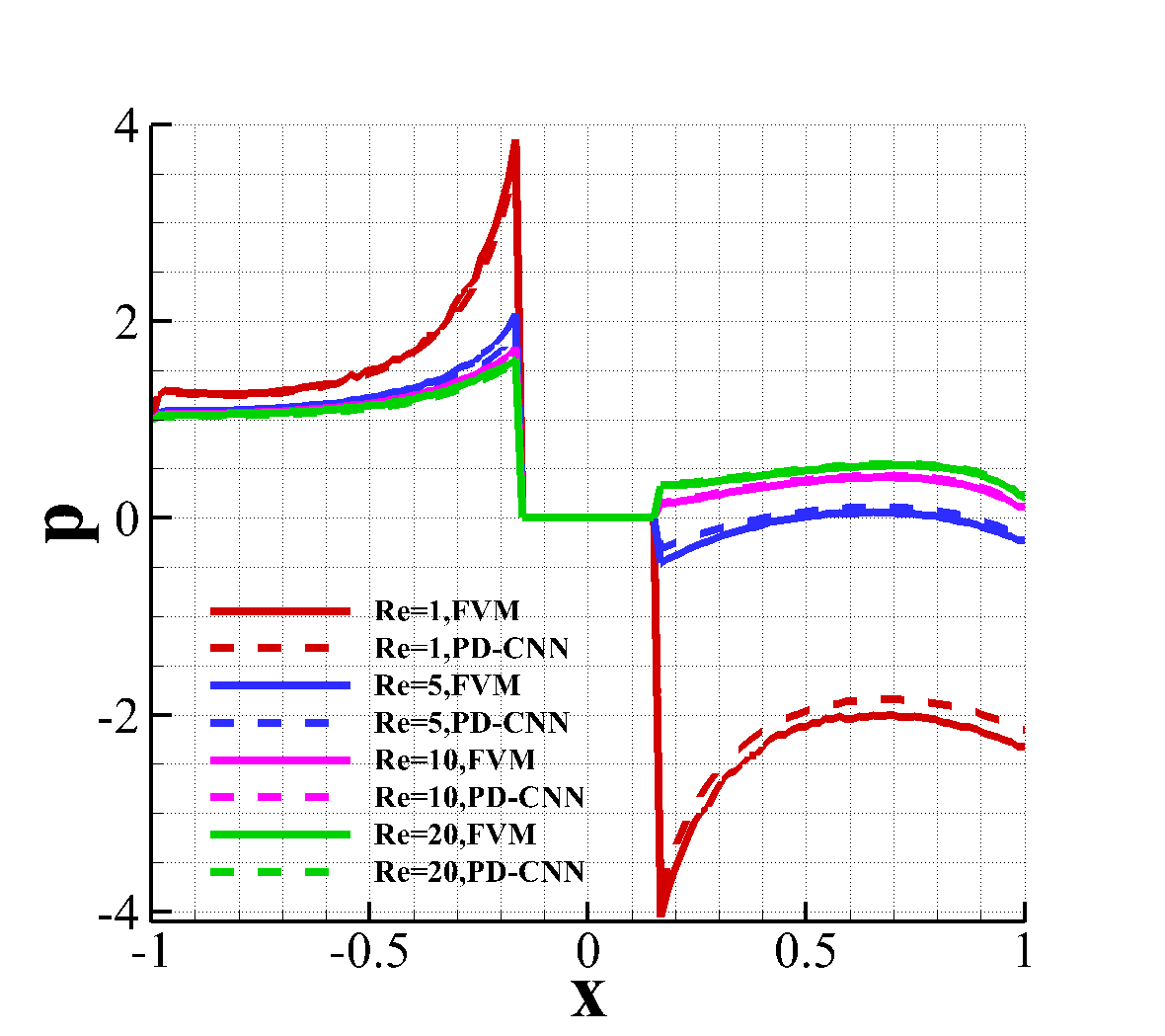}}
	\quad
	\subfloat[$u$]{
		\includegraphics[width=0.5\textwidth]{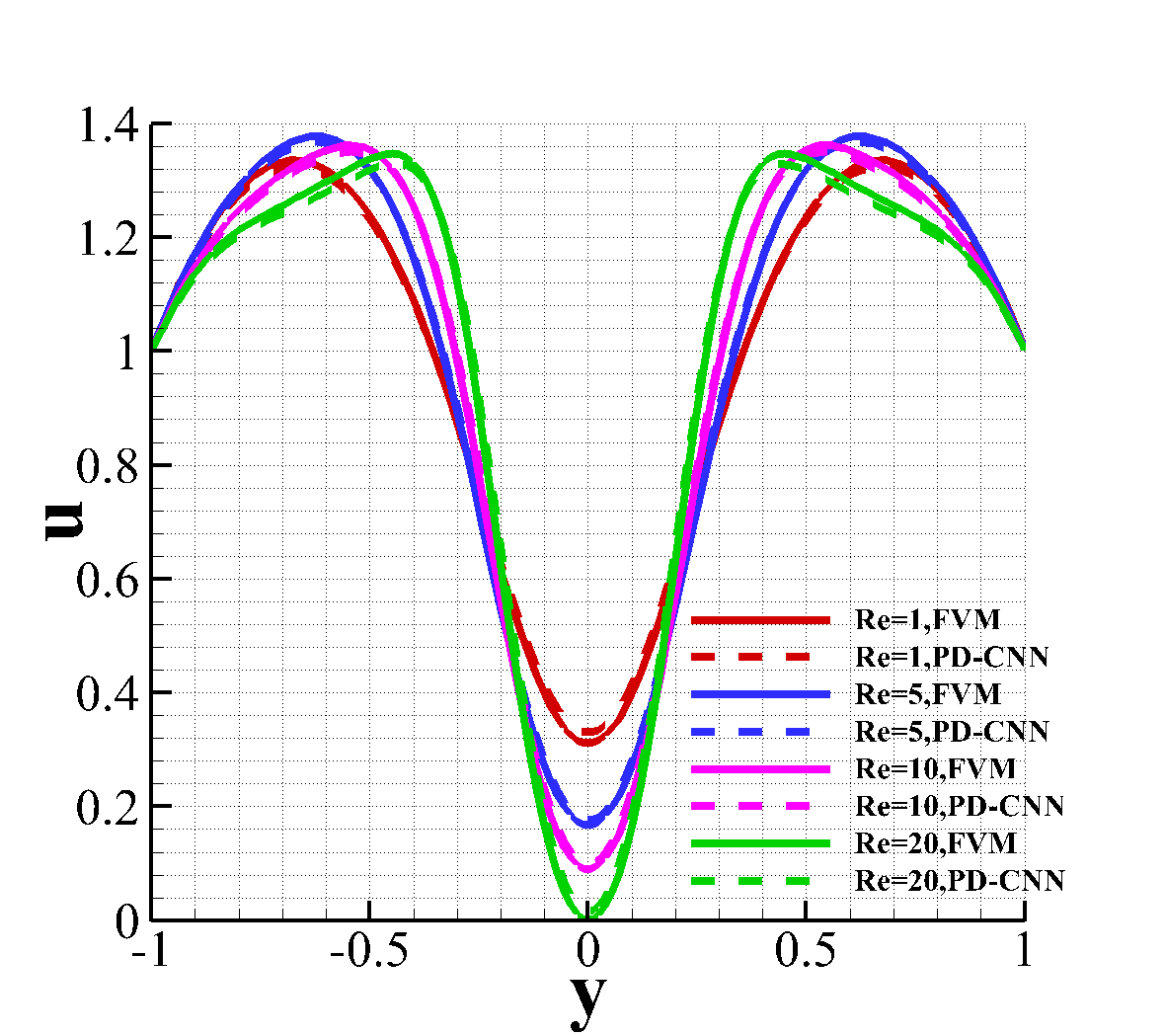}}
	\subfloat[$v$]{
		\includegraphics[width=0.5\textwidth]{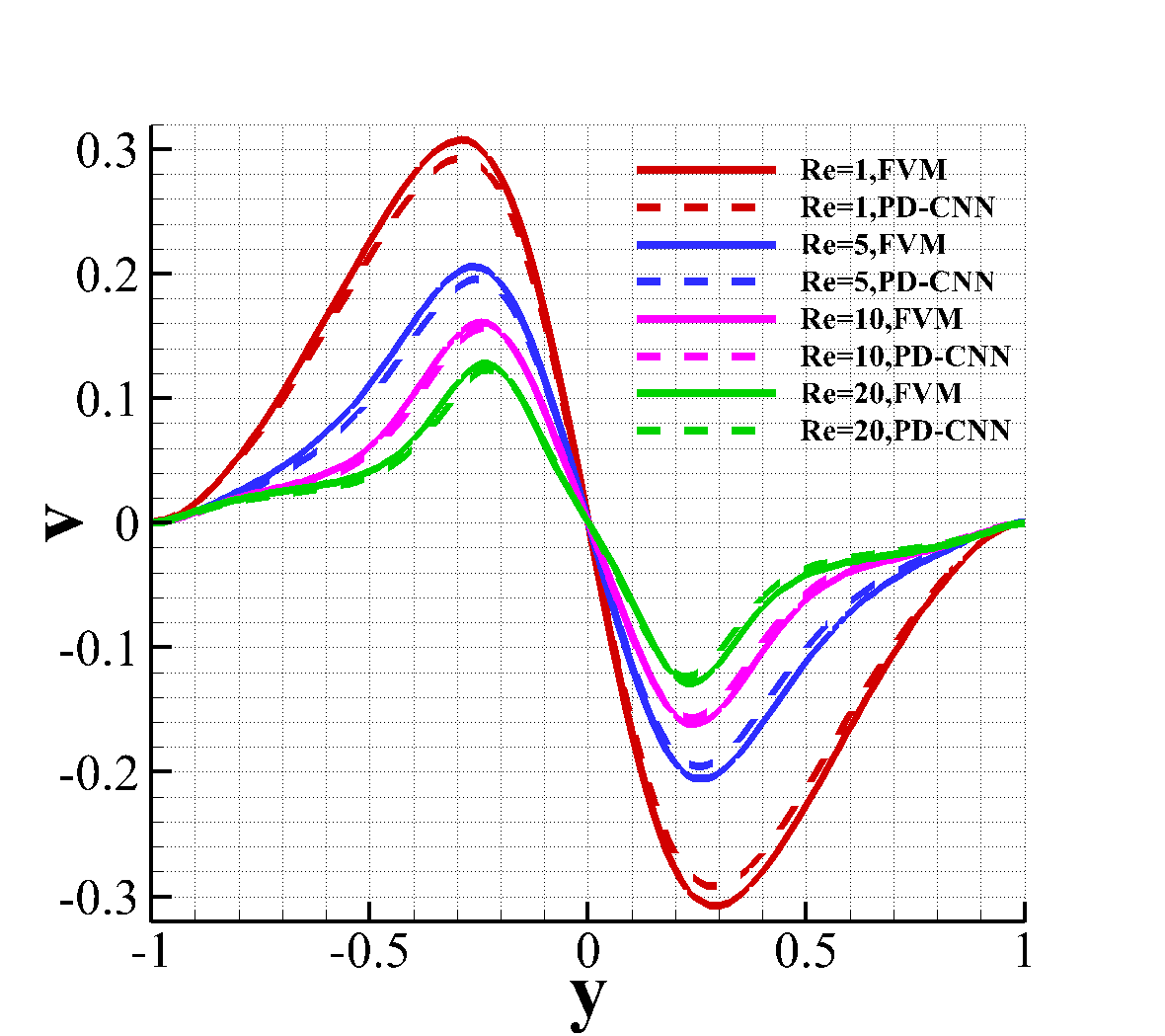}}
	\caption{Comparison between PD-CNN and FVM. The pressure is about $l_{1}:x=0.5$ and the velocities are about $l_{2}:y=0$. The results of lower $Re$ have relatively larger deviations.}
	\label{figs:322-Comparison}
\end{figure}

Figure \ref{figs:322-Comparison} shows the comparison between the learning results and the numerical solutions.
All the learning results of these four cases are in good agreement with the FVM results.
But the values have an ``even-bias". Compared with the ground truth, all values are closer to the mean value in the whole domain. 
For $p$, the mean value is in the vicinity of 0. For $u$, it is 0.7, and $v$ is 0. 
That means the generator has a trend to predict all the flow fields in the whole parameter space more evenly. 
In addition, the predictive flow fields which have lower Reynolds number have bigger deviations. 
Choosing $p$ as an example, the flow fields with relatively big Reynolds number have better accuracy, while the case of $Re=1$ has a bigger deviation.
In the generation of the training cases, the Reynolds numbers are chosen randomly between 0.8 and 20.2, which means the distribution of training cases is even.
However, for the case with $Re=1$, the flow field is more distinctive and requires more training effort.
So, when designing the training space, more training cases should be distributed near these exceptional parameter combinations.

%%%%%%%%%%%%%%%%%%%%%%%%%%%%%%%%%%%%%%%%%%%%%%%%%%%%%%%%%%%%%%%%%%%%%%%%%%%%%%%%
\section{Discussion}\label{sec:Discussion}
%%%%%%%%%%%%%%%%%%%%%%%%%%%%%%%%%%%%%%%%%%%%%%%%%%%%%%%%%%%%%%%%%%%%%%%%%%%%%%%%

\subsection{Accuracy}

The drag and lift coefficient of a cylinder can be expressed as follows \cite{schafer1996benchmark}
\begin{equation}\label{con:DragandLift}
C_{\rm d}=\frac{2F_{\rm D}}{\rho U^{2}A},\ 
C_{\rm l}=\frac{2F_{\rm L}}{\rho U^{2}A},
\end{equation}
where $A$ represents the frontal area. $F_{\rm D}$ and $F_{\rm L}$ denote the drag and lift force respectively.
When the mean physical residual $E$ of 1k epochs is lower than $0.4\%$, the training stops and the coefficients are calculated. 
These two coefficients obtained by PD-CNN are shown in Table 2.
For $C_{\rm d}$, PD-CNN with the resolution $384\times384$ predicts achieve a relative error  less than $2.9\%$.
The value of $C_{\rm l}$ is very small and the learning results are in the same magnitude compared with FVM references.
At the same time, it can be clearly seen that with the resolution increasing, both the drag and lift coefficients gradually approach the numerical solutions, which proves that the field solutions obtained by PD-CNN are convergent. 

In addition, we define the order of convergence as
\begin{equation}\label{con:OrderDefnition}
Order=\lvert \frac{{\rm log}({ e_i}/{ e_j })}{{\rm log}({h_i}/{h_j})} \rvert,
\end{equation}
where $e$ is the error of drag or lift coefficient under different resolutions, and $h$ is the resolution size. 
Based on this definition the PD-CNN can achieve approximate first-order accuracy.

\begin{table}[!h]
	\centering
	\caption{Accuracy of PD-CNN with different resolutions}
	\begin{tabular}{c|c|ccc|ccc}
		\toprule
		\toprule
		
		\multicolumn{1}{c|}{\multirow{2}[2]{*}{Coefficient}} &{\multirow{2}[2]{*}{Resolution}}&   &  $Re=1$ &   &   & $Re=20$ &  \\
		\multicolumn{1}{c|}{} &  & Result  & Error & Order & Result & Error & Order \\
		\midrule		
		
		\multicolumn{1}{c|}{\multirow{4}[2]{*}{$C_{\rm d}$}} & FVM & 
		44.3&
		- &
		-&
		5.96 &	
		-&
		-\\		
		%\cmidrule{2-8}
		
		\multicolumn{1}{c|}{ } & $128\times128$ &
		39.7&
		-4.6 &
		-&
		5.52 &	
		-0.44&
		-\\		
		%\cmidrule{2-8}
		
		\multicolumn{1}{c|}{ } & $256\times256$ & 
		42.3&
		-2.0 &
		1.21&
		5.71 &	
		-0.25&
		0.82\\		
		%\cmidrule{2-8}
		
		\multicolumn{1}{c|}{ } & $384\times384$ &
		43.0&
		-1.3 &
		1.08&
		5.89 &	
		-0.07&
		3.09\\					  		
		\midrule
		
		\multicolumn{1}{c|}{\multirow{4}[2]{*}{$C_{\rm l}$}} & FVM & 
		0.0028&
		- &
		-&
		0.0097 &	
		-&
		-\\		
		%\cmidrule{2-8}
		
		\multicolumn{1}{c|}{ } & $128\times128$ &
		-0.059&
		-0.0618 &
		-&
		-0.044 &	
		-0.0537&
		-\\		
		%\cmidrule{2-8}
		
		\multicolumn{1}{c|}{ } & $256\times256$ & 
		-0.015&
		-0.0178 &
		1.79&
		-0.013 &	
		-0.0227&
		1.24\\		
		%\cmidrule{2-8}
		
		\multicolumn{1}{c|}{ } & $384\times384$ &
		-0.0059&
		-0.0087 &
		1.76&
		0.0028 &	
		-0.0069&
		2.94\\		

		\bottomrule
		\bottomrule
	\end{tabular}%
	\label{tab:steps}%
\end{table}%

\subsection{Acceleration}
In the numerical experiments of PD-CNN which are accelerated with reference targets, there are 18 cases in each batch. 
The first half batch consists of 9 random $Re$ varied with epoch number, while the other half batch consists of 9 constant $Re$ fixed in the whole training process.    
According to Equation (\ref{con:weightedLossNew}), a moderate number of constant $Re$ for the reference targets requires being defined beforehand.
As discussed in section \ref{sec:MultipleCases}, the cases whose Reynolds numbers are near to 1 are much more difficult to train. 
So, the manually defined 9 constant Reynolds numbers are 1.0, 1.5, 2.0, 3.0, 4.0, 6.0, 8.0, 12.0 and 18.0.
In the whole training process, the numerical solutions obtained by FVM of these 9 constant Reynolds numbers are input as nine targets and the CNN is trained to minimize Equation (\ref{con:LossData}).

\begin{figure}[!h]
	\centering
	\subfloat[FVM]{
		\includegraphics[width=0.33\textwidth]{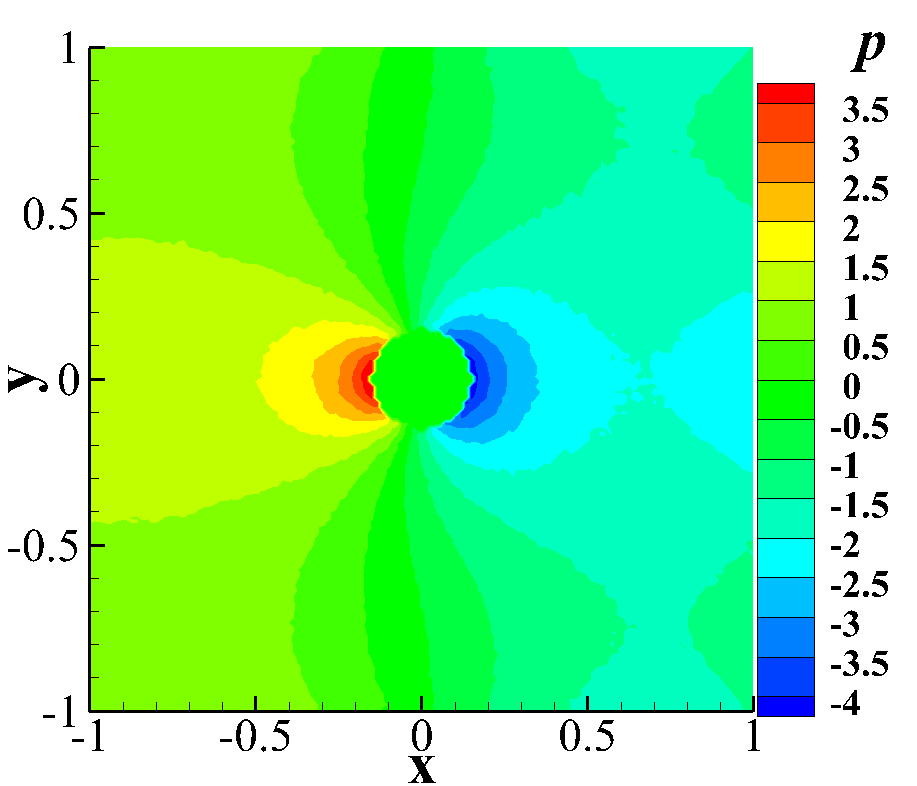}}
	\subfloat[PD-CNN]{
		\includegraphics[width=0.33\textwidth]{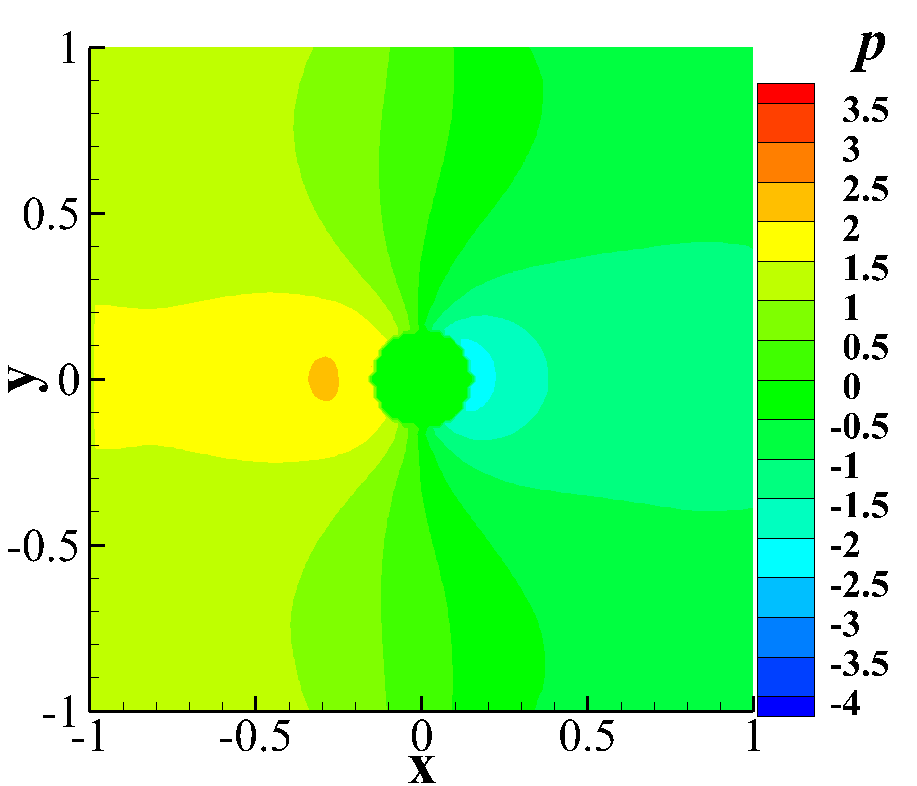}}
	\subfloat[PD-CNN with references]{
		\includegraphics[width=0.33\textwidth]{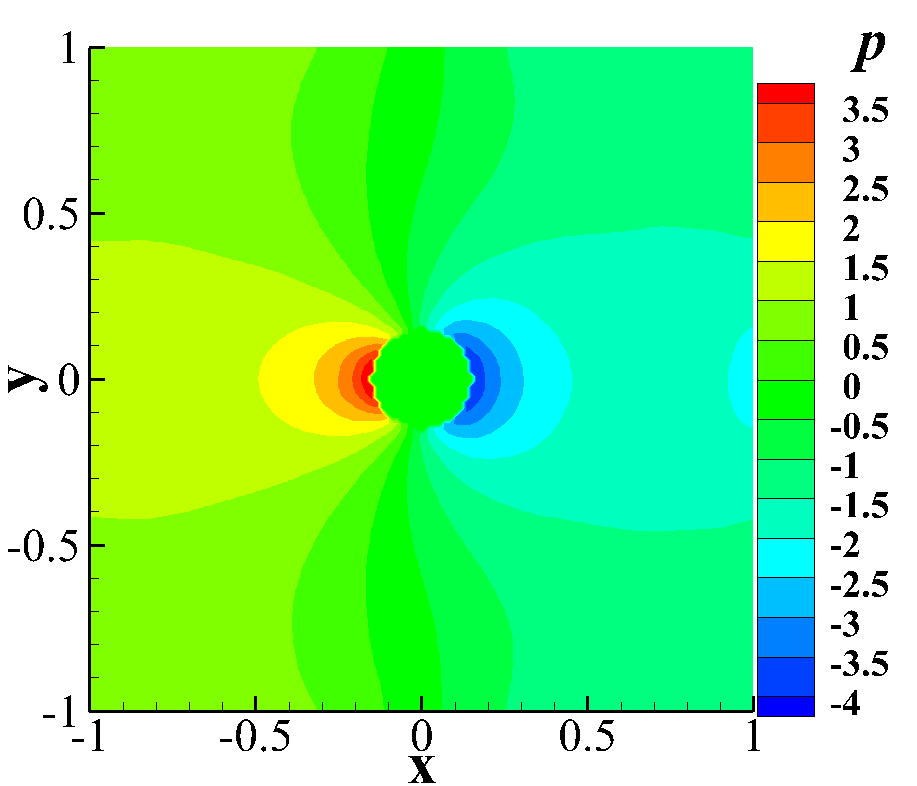}}
	\caption{Comparison of pressure field of different methods when $Re=1$ (epoch=10k). For this typical case with reference, the training with targets is able to obtain accurate solution faster.}
	\label{figs:411-P}
\end{figure}

\begin{figure}[!h]
	\centering
	\subfloat[FVM]{
		\includegraphics[width=0.33\textwidth]{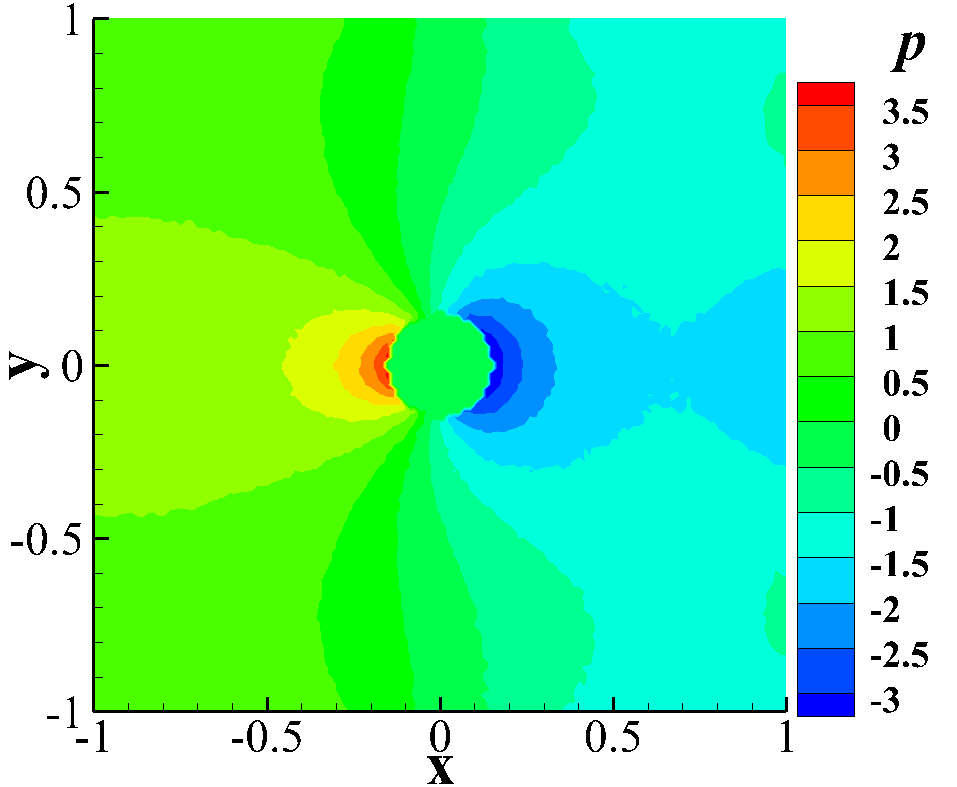}}
	\subfloat[PD-CNN]{
		\includegraphics[width=0.33\textwidth]{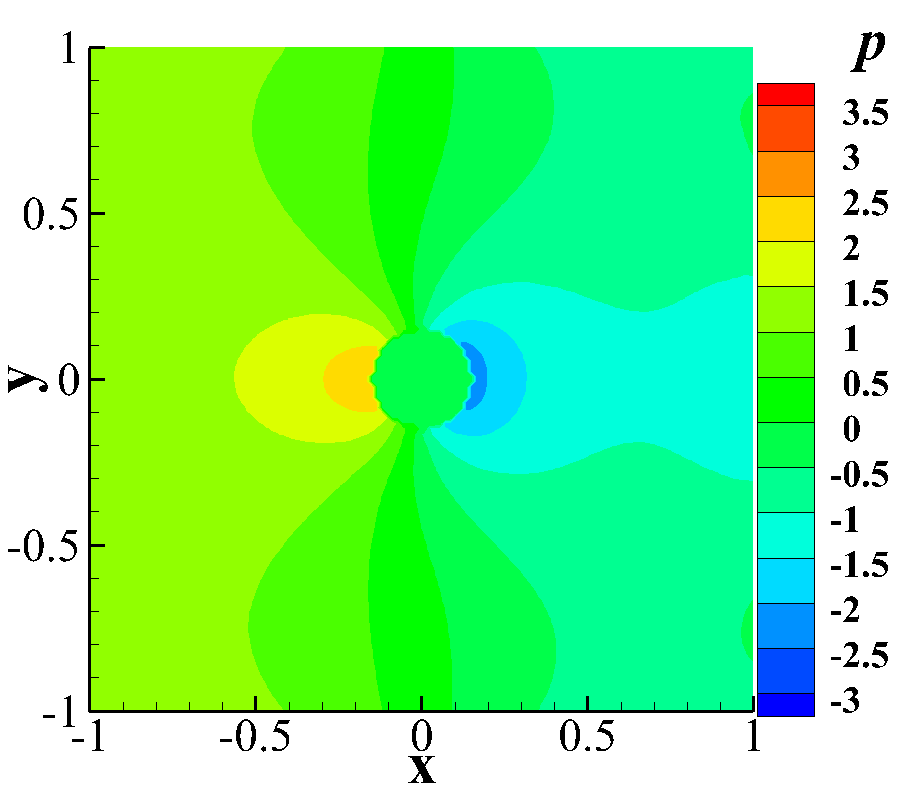}}
	\subfloat[PD-CNN with references]{
		\includegraphics[width=0.33\textwidth]{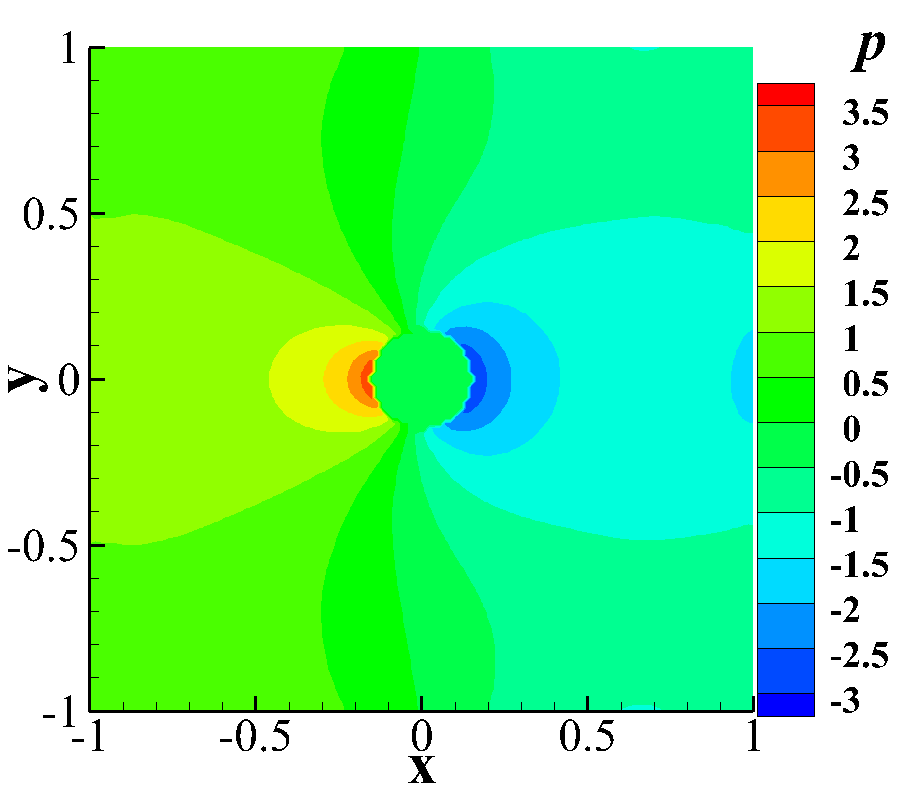}}
	\caption{Comparison of pressure field of different methods when $Re=1.2$ (epoch=10k). For this typical case without reference, the training with targets is also able to obtain accurate solution faster.}
	\label{figs:412-P}
\end{figure}

In this way, the reference targets restrain the results generated by the physics-driven method to approximate real solutions faster (as shown in Figure \ref{figs:411-P} and Figure \ref{figs:412-P}).
Since the targets only include a limiting number of cases and are used through the whole training process, this approach reduces the expensive data generation cost of the traditional data-driven methods.
In practical engineering applications, the reference targets can be easily picked from the existing data, e.g., experimental and numerical results.

%%%%%%%%%%%%%%%%%%%%%%%%%%%%%%%%%%%%%%%%%%%%%%%%%%%%%%%%%%%%%%%%%%%%%%%%%%%%%%%%
\section{Conclusion}\label{sec:Conclusion}
%%%%%%%%%%%%%%%%%%%%%%%%%%%%%%%%%%%%%%%%%%%%%%%%%%%%%%%%%%%%%%%%%%%%%%%%%%%%%%%%

In this paper, we proposed a physics-driven method based on a CNN with a U-net structure.
By introducing the discretized Navier-Stokes equations and boundary conditions as the loss function, the CNN is able to directly predict steady-state laminar flow fields.
Compared with the MLP used in previous research, the CNN is able to embed the objective geometry and flow structure into the latent space, and then decode them to reconstruct the corresponding flow fields with physics consistency.
The PD-CNN is able to obtain accurate solutions for both single case and multiple cases.
In the multiple cases prediction of flow around cylinder, PD-CNN is capable of describing the transformation of the ``twin-vortices" and obtaining the critical Reynolds number between creeping laminar state and laminar flow with separation.
By constraining with a small number of reference targets, the network training was accelerated, especially for the difficult cases.
Further research will be carried out for predicting complex
flow fields in practical engineering problems with the present method.

%%%%%%%%%%%%%%%%%%%%%%%%%%%%%%%%%%%%%%%%%%%%%%%%%%%%%%%%%%%%%%%%%%%%%%%%%%%%%%%%
\section{Acknowledgement}
%%%%%%%%%%%%%%%%%%%%%%%%%%%%%%%%%%%%%%%%%%%%%%%%%%%%%%%%%%%%%%%%%%%%%%%%%%%%%%%%

Hao Ma (No. 201703170250) and Yuxuan Zhang (No. 201804980021) are supported by China Scholarship Council when they conduct the work this paper represents.

\clearpage

\bibliography{mybibfile3}

\end{document}